\documentclass[aip,reprint,jcp]{revtex4-1}
\usepackage[utf8]{inputenc}
\usepackage[T1]{fontenc}
\usepackage{lmodern}

\usepackage{graphicx}
\usepackage{amsmath}
\usepackage{amsfonts}
\usepackage{amssymb}
\usepackage{fixmath}
\usepackage{microtype}
\usepackage{hyperref}
\usepackage{braket}
\usepackage{multirow}
\newcommand{\specialcell}[2][c]{ %
  \begin{tabular}[#1]{@{}c@{}}#2\end{tabular}}

\newcommand*{\abs}[1]{\left|#1\right|}
\newcommand*{\pvec}[1]{\vec{#1}\mkern2mu\vphantom{#1}'}
\usepackage{bm}\newcommand*{\mat}[1]{\bm{#1}}
\newcommand*{\spinup}{\uparrow}
\newcommand*{\spindown}{\downarrow}
\newcommand*{\dd}{\mathrm{d}}

\begin{document}

\title{Tight-Binding Approximations to Time-Dependent Density Functional Theory\\--- a fast approach for the calculation of electronically excited states} 


\author{Robert R\"uger}
\email{rueger@scm.com}
\affiliation{Scientific Computing \& Modelling NV, De Boelelaan 1083, 1081 HV Amsterdam, The Netherlands}
\affiliation{Department of Theoretical Chemistry, Vrije Universiteit Amsterdam, De Boelelaan 1083, 1081 HV Amsterdam, The Netherlands}
\affiliation{Wilhelm-Ostwald-Institut für Physikalische und Theoretische Chemie, Linnéstr. 2, 04103 Leipzig, Germany}

\author{Erik van Lenthe}
\affiliation{Scientific Computing \& Modelling NV, De Boelelaan 1083, 1081 HV Amsterdam, The Netherlands}

\author{Thomas Heine}
\affiliation{Wilhelm-Ostwald-Institut für Physikalische und Theoretische Chemie, Linnéstr. 2, 04103 Leipzig, Germany}

\author{Lucas Visscher}
\affiliation{Department of Theoretical Chemistry, Vrije Universiteit Amsterdam, De Boelelaan 1083, 1081 HV Amsterdam, The Netherlands}

\date{\today}

\begin{abstract}
We propose a new method of calculating electronically excited states that combines a density functional theory~(DFT) based ground state calculation with a linear response treatment that employs approximations used in the time-dependent density functional based tight binding~(TD-DFTB) approach.
The new method termed TD-DFT+TB does not rely on the DFTB parametrization and is therefore applicable to systems involving all combinations of elements.
We show that the new method yields UV/Vis absorption spectra that are in excellent agreement with computationally much more expensive time-dependent density functional theory~(TD-DFT) calculations.
Errors in vertical excitation energies are reduced by a factor of two compared to TD-DFTB.
\end{abstract}

\pacs{31.15.ee}

\maketitle

\section{Introduction}

Owing its success to the good compromise between accuracy and computational cost, density functional theory (DFT) based on the theorems~\cite{HohenbergKohnTheorem1964} by \citeauthor{HohenbergKohnTheorem1964} and employing the Kohn-Sham ansatz~\cite{KohnShamEquations1965} for the kinetic energy has become the most widely used method in both quantum chemistry and solid state theory over the last few decades.
Rooted in the Kohn-Sham DFT framework, density functional based tight binding (DFTB)~\cite{PorezagDFTB1995,SeifertDFTB1996} has been developed as a computationally very efficient approximation to DFT for systems too large to be treated with its parent method.
DFTB's efficiency stems from the use of an optimized minimum valence orbital basis that reduces the linear algebra operations, and a two center-approximation for the Kohn-Sham potential that allows precalculation and storage of integrals using the Slater-Koster technique~\cite{SlaterKoster1954}.
The self-consistent charge extension (SCC-DFTB, recently also called DFTB2)~\cite{SeifertSCCDFTB1998} accounts for density fluctuations and improves results for systems with polar bonds.
A further extension known as DFTB3~\cite{ElstnerDFTB32011} has been developed to improve the description of hydrogen-bonded complexes and proton affinities.
While DFTB is much more efficient than DFT it requires careful parametrization for all involved elements in order to yield accurate results.
The limited availability of these parameters has historically slowed down the adoption of DFTB, but general purpose parameter sets covering large parts of the periodic table have recently become available~\cite{Elstner3obParameters2013,Elstner3obSPParameters2014,Elstner3obMgZnParameters2015,Elstner3obMoreParameters2015,WahiduzzamanQuasinano2013,OliveiraQUASINANO2015}.

As the Hohenberg-Kohn theorems only concern the ground state of a system, density functional theory is not applicable to the broad class of problems involving electronically excited states.
The foundation for the excited state extension known as time-dependent density functional theory (TD-DFT) was later laid by \citeauthor{RungeGrossTheorem1984}, who generalized Hohenberg and Kohn's theorems to time-dependent external potentials~\cite{RungeGrossTheorem1984}.
Based on their work, \citeauthor{CasidaTDDFT1995} calculated the linear response of the electron density to a perturbation in the external potential and from this derived an eigenvalue equation in the space of single orbital transitions from which the excited states of the electronic system can be obtained~\cite{CasidaTDDFT1995}.
In the field of quantum chemistry, \citeauthor{CasidaTDDFT1995}'s TD-DFT approach is today probably the most widely used method for the calculation of excited state properties.
A recent review of TD-DFT can be found in reference~\citenum{CasidaTDDFTReview2012}.

An excited state calculation using TD-DFT is computationally quite demanding, much more so than the underlying DFT calculation of the ground state.
For many systems it is therefore feasible to calculate the ground state, while a calculation of excited states is computationally out of reach.
Various ways to reduce the computational complexity of TD-DFT have been put forward; based for example on partitioning into subsystems~\cite{CasidaSubsysTDDFT2003,NeugebauerSubsysTDDFT2007}, neglect of terms~\cite{HirataTDA1999,GrimmeSimplifiedTDA2013,GrimmeRangeSepsTDA2014}, truncation of the single orbital transition space~\cite{GrimmeSimplifiedTDA2013,RugerIntensitySelection2015} and approximation of integrals~\cite{GrimmeSimplifiedTDA2013,GrimmeRangeSepsTDA2014,GrimmeSimplifiedTDDFT2014,NiehausTDDFTB2001,NiehausTDDFTBOnsiteAndFracOcc2013,GaussianTDDFTB2011}.
Among the methods that approximate integrals is time-dependent density functional theory based tight binding (TD-DFTB)~\cite{NiehausTDDFTB2001,NiehausTDDFTBOnsiteAndFracOcc2013,GaussianTDDFTB2011} developed by \citeauthor{NiehausTDDFTB2001}, which builds on a DFTB ground state calculation and translates \citeauthor{CasidaTDDFT1995}'s linear response approach to the framework of DFTB.
It has been found to yield very good results for $\pi \rightarrow \pi^*$ transitions~\cite{NiehausTDDFTB2001}, making it especially suitable for the calculation of UV/Vis absorption spectra~\cite{RugerIntensitySelection2015}.
Being computationally much cheaper than a full TD-DFT calculation and applicable to very large systems, TD-DFTB has been used in a variety of applications~\cite{tddftbapp1_doi:10.1021/jp026752s,tddftbapp2_PhysRevB.73.205312, tddftbapp3_doi:10.1021/jp071125u,tddftbapp4_doi:10.1021/ct700041v,tddftbapp5_doi:10.1021/jp065704v,tddftbapp6_10.1063/1.2715101,tddftbapp7_doi:10.1063/1.2940735,BonacicNonAdMDWithTDDFTB2009,tddftbapp9_PSSB:PSSB201100719,tddftbapp10_Fan201417} in which TD-DFT would not have been feasible.
A review of TD-DFTB can be found in reference~\citenum{NiehausTDDFTBReview2009}.\looseness -1

TD-DFTB inherits certain limitations from the DFTB ground state calculation it is based on:
The electronic structure from DFTB, which is the basis of the excited state calculation, is of limited accuracy compared to a DFT calculation with a reasonable choice of exchange-correlation functional and orbital basis.
Furthermore, whereas TD-DFT can be used for any system, historically the applicability of TD-DFTB was restricted to systems involving only elements for which DFTB parameters are available. With the development of the QUASINANO parameters, which are available throughout the periodic table, by \citeauthor{WahiduzzamanQuasinano2013}~\cite{WahiduzzamanQuasinano2013} this drawback was removed for the electronic part (and thus for TD-DFTB), even though TD-DFTB still requires a careful performance validation for the target system class.

These limitations are insofar particularly unfortunate as it is neither the calculation of the ground state's electronic structure that is the computational bottleneck, nor requires the application of tight-binding approximations to the TD-DFT concept within Casida's formulation any parameterization effort.
In this article we introduce TD-DFT+TB, a new method for calculating electronically excited states that combines a DFT ground state with a linear response treatment that employs approximations similar to the ones used in TD-DFTB.
We show that the cost of this calculation is approximately the same as of a ground state DFT calculation, and the accuracy of the excited state properties is much better than TD-DFTB.

The approach proposed in this article is inspired by and closely related to the sTDA~\cite{GrimmeSimplifiedTDA2013,GrimmeRangeSepsTDA2014} and sTD-DFT~\cite{GrimmeSimplifiedTDDFT2014} methods developed by~\citeauthor{GrimmeSimplifiedTDA2013}~et al., which also use a DFT ground state calculation and make TD-DFTB like approximations in \citeauthor{CasidaTDDFT1995}'s formalism.
The main difference compared to our approach is, that these methods are based on DFT with a hybrid exchange-correlation functional~\cite{BeckeHybrids1993}.
Hybrid functionals are usually employed to correct the underestimated charge-transfer excitation energies in TD-DFT with local functionals.
However, we argue that for the calculation of optical absorption spectra, underestimated charge-transfer excitation energies only constitute a technical problem that can be solved conceptually easier and computationally more efficient by employing a physically motivated truncation of the single orbital transition space~\cite{RugerIntensitySelection2015}.
Furthermore, it was recently shown~\cite{BaerendsKSOrbitals2013,BaerendsKSOrbitals2014} by~\citeauthor{BaerendsKSOrbitals2013} that excitations loose their single orbital transition character with the admixture of Hartree-Fock exact exchange, which complicates the interpretation of the results.
We therefore believe that the calculation of excited states of large systems should also be approached from a pure density functional standpoint.
Both sTDA and sTD-DFT have been developed for hybrid functionals and contain free parameters that have been fitted to yield good results between 20\% and 60\% exact exchange.
As such, they are not intended for and can not directly be used with local functionals.
We believe that TD-DFT+TB as proposed in this article complements sTDA and sTD-DFT by making approximate TD-DFT methods also available for local exchange-correlation functionals.

The remainder of this article is organized as follows.
In section~\ref{s:method_review} we recapitulate the most important equations from DFT and DFTB as well as their linear response extensions in order to set the stage for section~\ref{s:TD-DFT+TB}, in which we motivate and introduce TD-DFT+TB.
We will also discuss its relation to other approximate TD-DFT methods, such as TD-DFTB, sTDA and sTD-DFT.
In section~\ref{s:results} we evaluate the accuracy and performance of the new method by calculating vertical excitation energies for a benchmark set of molecules as well as the UV/Vis absorption spectra of selected compounds.
Section~\ref{s:conclusion} summarizes our results and concludes the article.

\section{\label{s:method_review}Review of methods}

In order to establish the notation for the remainder of this article, this section contains a short summary of DFT and DFTB as well as their linear response extensions.

\subsection{Molecular orbitals from DFT(B)}

Electronic structure calculations of molecular systems typically use atom centered basis functions~$\chi_\mu(\vec r)$, so that the molecular orbitals~$\phi_i(\vec r)$ can be written as
\begin{equation}\label{eq:OrbitalExpansion}
\phi_i(\vec r) = \sum_{\mathcal A}^{N_\mathrm{atom}} \sum_{\mu \in \mathcal A} c_{\mu i} \chi_\mu(\vec r) \; .
\end{equation}
The basis functions $\chi_\mu(\vec r)$ are composed of primitives that may be Gaussian, Slater, numerical or any other functions that are centered at the atomic positions.
For DFT the size of the basis is variable and can within the limits of computational affordability be chosen according to the desired accuracy, while DFTB on the other hand typically uses an optimized minimum valence orbital basis that is fixed during the DFTB parameter creation, and can not be changed at run-time.

The expansion coefficients~$c_{\mu i}$ of the molecular orbitals are obtained by solving the secular equations
\begin{equation}\label{eq:KohnShamSecularEquation}
\sum_\nu H_{\mu\nu} c_{\nu i} = \varepsilon_i \sum_\nu S_{\mu \nu} c_{\nu i} \; .
\end{equation}
Here,
\begin{equation}
S_{\mu \nu} = \int \dd \vec r \; \chi_\mu(\vec r) \chi_\nu(\vec r)
\end{equation}
is the overlap between basis functions.
In DFT, the Hamiltonian matrix elements~$H_{\mu \nu}$ are calculated as
\begin{equation}
H_{\mu \nu} = \int \dd \vec r \; \chi_\mu(\vec r) \left( - \frac{1}{2} \nabla^2 + v_\text{eff}(\vec r) \right) \chi_\nu(\vec r) \; ,
\end{equation}
where
\begin{equation}
v_\text{eff}(\vec r) = v_\text{ext}(\vec r) + \int \dd \pvec r \frac{\rho(\pvec r)}{\abs{\vec r - \pvec r}} + \frac{\delta E_\mathrm{xc}[\rho]}{\delta \rho(\vec r)}
\end{equation}
is the Kohn-Sham effective potential~\cite{KohnShamEquations1965}, consisting of the external potential, an electrostatic term and the so-called exchange-correlation potential.
Note that the effective potential~$v_\text{eff}(\vec r)$ depends on the molecular orbitals themselves through their electronic density~$\rho(\vec r)$, so that equation~\eqref{eq:KohnShamSecularEquation} has to be solved self-consistently.

DFTB avoids the evaluation of integrals at run-time by replacing the actual density $\rho$ by a trial density $\rho_0$.
This trial density is a superposition of atomic contributions which are optimized within the parameterization process.
Within the DFTB-inherent two-center approximation the effective potential is constructed by superposing two spherical atomic effective potentials~\cite{PorezagDFTB1995,SeifertDFTB1996,WahiduzzamanQuasinano2013} or trial densities~\cite{SeifertSCCDFTB1998,ElstnerDFTB32011}, which allows the Hamiltonian and overlap matrix elements to be precalculated and stored using the Slater-Koster technique~\cite{SlaterKoster1954}.
This shifts the computational bottleneck from the calculation of matrix elements to linear algebra operations which are dominated by the diagonalization.
Together with the small matrix sizes due to the minimal valence basis set, DFTB is computationally extremely efficient.

\subsection{Excited states and absorption spectra from TD-DFT}

Once the electronic structure of the ground state has been determined, excited states can be calculated using Casida's linear response approach~\cite{CasidaTDDFT1995}, which casts the problem of calculating excitation energies and excited states into an eigenvalue equation in the~$N_\text{trans} = N_\text{occ} N_\text{virt}$ dimensional space of single orbital transitions.
The eigenvalue problem can be written as
\begin{equation}\label{eq:CasidasEquation}
\sum_{jk} \Omega_{ia,jb} F_{jb,I} = \Delta_I^2 F_{ia,I} \; ,
\end{equation}
where $\Delta_I$ is the vertical excitation energy of the $I$-th excited state. We adopt the usual convention of using the indexes~$i,j$ for occupied and~$a,b$ for virtual orbitals.
The elements~$F_{ia,I}$ correspond to the contribution of the transition from the occupied orbital~$\phi_i$ to the virtual orbital~$\phi_a$
and can be used to construct an approximate excited state wavefunction~$\ket{\psi}$ from the Slater determinant~$\ket{\psi_0}$ of the occupied Kohn-Sham orbitals~\cite{CasidaTDDFT1995}.
\begin{equation}\label{eq:CasidaAssignmentAnsatz}
\ket{\psi_I} = \sum_{ia} \sqrt{\frac{2\Delta_{ia}}{\Delta_I}} F_{ia,I} \; \hat c^\dagger_a \hat c^{\phantom\dagger}_i \ket{\psi_0}
\end{equation}
Here we use~$\Delta_{ia} = \varepsilon_a - \varepsilon_i$ for the difference in orbital energy between the involved Kohn-Sham orbitals.
The elements of the matrix~$\mat \Omega$ are given by
\begin{equation}\label{eq:CasidaMatrixElements}
\Omega_{ia,jb} = \delta_{ij} \delta_{ab} \Delta_{ia}^2  + 4 \sqrt{\Delta_{ia}\Delta_{jb}} K_{ia,jb} \; ,
\end{equation}
and looking back at equation~\eqref{eq:CasidasEquation}, it is easy to see that it is the so-called coupling matrix~$\mat K$ that shifts the excitation energies~$\Delta_I$ away from the orbital energy differences~$\Delta_{ia}$.
The coupling matrix depends on the multiplicity of the calculated excitations.
For the sake of clarity we will restrict our discussion to singlet excitations for the moment.
Triplet excitations pose no additional problems and their calculation will be discussed later.
For singlet excitations in TD-DFT the elements of the coupling matrix are given by
\begin{align}\label{eq:CouplingMatrix_TDDFT}
K^\mathrm{S}_{ia,jb} &= \int \dd^3 \vec r \int \dd^3 \pvec r \phi_i(\vec r) \phi_a(\vec r) \times \\
&  \hspace{52.8pt} \times f_\mathrm{Hxc}[\rho^\mathrm{GS}](\vec r, \pvec r) \; \phi_j(\pvec r) \phi_b(\pvec r) \; ,\nonumber
\end{align}
where the kernel
\begin{equation}\label{eq:Hxc_Kernel}
f_\mathrm{Hxc}[\rho^\mathrm{GS}](\vec r, \pvec r) = \frac{1}{\abs{\vec r - \pvec r}} + \frac{\delta^2 E_\mathrm{xc}[\rho]}{\delta \rho(\vec r) \delta \rho(\pvec r)} \Big|_{\rho^\mathrm{GS}}
\end{equation}
incorporates both a Coulomb term and the second derivative of the DFT exchange-correlation functional~$E_\mathrm{xc}[\rho]$.

For the prediction of photon absorption spectra it is necessary to calculate both the excitation energies~$\Delta_I$ as well as the corresponding transition dipole moments~$\vec d_I$.
Excitation energies are immediately obtained as the eigenvalues of Casida's equation~\eqref{eq:CasidasEquation}, and using the eigenvector elements~$F_{ia,I}$ the transition dipole moments~$\vec d_I$ can be calculated as a linear combination of the transition dipole moments~$\vec d_{ia}$ of the single orbital transitions.
\begin{equation}
\label{eq:TDM} \vec d_I = \sum_{ia} \sqrt{\frac{2\Delta_{ia}}{\Delta_I}} F_{ia,I} \, \vec d_{ia}
\end{equation}
Here the transition dipole moments~$\vec d_{ia}$ of the single orbital transitions are calculated as
\begin{equation}
\label{eq:SOTDM_exact} \vec d_{ia} = \int \dd^3 \vec r \; \phi_i(\vec r) \phi_a(\vec r) \, \vec r  \; .
\end{equation}
In order to make a connection to experimentally measured quantities, the theoretically calculated oscillator strengths~$f_I$ and excitation energies~$E_I$ can be related~\cite{MullikenIntensities1939} to the molar absorptivity
\begin{equation}
\epsilon(E) = \frac{\pi}{2 \ln(10)} \frac{N_A e^2 \hbar}{m_e c \varepsilon_0} \sum_I f_I \, \Gamma(E - E_I) \; .
\end{equation}
Here $\Gamma(E)$ is a normalized, typically peaked function that models the experimental line broadening. Both Gaussian and Lorentzian functions are common choices for~$\Gamma(E)$.

It would be beyond the scope of this article to go into further details on the properties and problems of TD-DFT.
A recent review of the strengths and weaknesses of TD-DFT in general can be found in reference~\citenum{CasidaTDDFTReview2012}.
There is also an excellent book on TD-DFT, see reference~\citenum{TDDFTBook2012}.
The method put forward in this article presents an approximation to TD-DFT and we therefore consider TD-DFT with a GGA exchange-correlation functional as the reference method.

\subsection{\label{ss:TDDFTB}TD-DFTB as an approximation to TD-DFT}

The calculation of the TD-DFT coupling matrix elements involves expensive two-center integrals and even though highly optimized implementations are available~\cite{GisbergenADFTDDFTB1999}, evaluating the integrals is still the computational bottleneck of the method.
In order to make density functional based excited state calculations applicable to larger systems, \citeauthor{NiehausTDDFTB2001} have put forward TD-DFTB~\cite{NiehausTDDFTB2001,NiehausTDDFTBOnsiteAndFracOcc2013}, which builds on top of a DFTB ground state calculation and uses DFTB-like approximations for the coupling matrix elements:
The transition density $\phi_i(\vec r) \phi_a(\vec r)$ in equation~\eqref{eq:CouplingMatrix_TDDFT} is subjected to a multipole expansion truncated at first order (monopole approximation)
\begin{equation}\label{eq:MonopoleApproximation_TDDFTB}
\phi_i(\vec r) \phi_a(\vec r) \approx \sum_\mathcal A q_{ia,\mathcal A} \, \xi_\mathcal A(\vec r) \; ,
\end{equation}
where $\xi_\mathcal A(\vec r)$ is a spherically symmetric function centered on atom~$\mathcal A$.
This allows the singlet-singlet coupling matrix elements in TD-DFTB to be written as
\begin{equation}\label{eq:CouplingMatrix_TDDFTB}
K^\mathrm{S}_{ia,jb} = \sum_\mathcal{AB} q_{ia,\mathcal A} \, \gamma_\mathcal{AB} \, q_{jb,\mathcal B} \; .
\end{equation}
The so-called atomic transition charges~$q_{ia,\mathcal A}$ are calculated from the molecular orbital coefficient and overlap matrix~$\mat C$ and~$\mat S$ using Mulliken population analysis~\cite{MullikenPopulationAnalysis1955}.
\begin{equation}\label{eq:AtomTransCharge_Mulliken}
q_{ia,\mathcal A} = \frac{1}{2} \sum_{\mu \in \mathcal A} \sum_{\nu} \Big( c_{\mu i} S_{\mu \nu} c_{\nu a} + c_{\nu i} S_{\nu \mu} c_{\mu a} \Big)
\end{equation}
While $\gamma_\mathcal{AB}$ should in principle be calculated as a two-center integral over the product of the atom centered functions $\xi_\mathcal {A/B}$ and the kernel $f_\mathrm{Hxc}$ from equation~\eqref{eq:Hxc_Kernel}, it is in practice approximated as a function
\begin{equation}
\gamma_\mathcal{AB} = \gamma_\mathcal{AB}\left(\eta_\mathcal{A}, \eta_\mathcal{B}, \abs{\vec R_\mathcal{A} - \vec R_\mathcal{B}}\right)
\end{equation}
of the internuclear distance and the chemical hardness~$\eta_\mathcal{A}$ and~$\eta_\mathcal{B}$ of atom~$\mathcal A$ and~$\mathcal B$ respectively, converging to the Coulomb interaction between two point charges for long distances~\cite{NiehausTDDFTB2001,SeifertSCCDFTB1998}.
The required atomic chemical hardness is not a free, tunable parameter, but rather an inherent property of the atoms themselves. There is, however, some freedom in the choice of the method used to obtain these values, e.g.\ from atomic DFT calculations by application of Janak's theorem~\cite{MinevaHeine1,MinevaHeine2,SeifertSCCDFTB1998}, or using a phenomenological model~\cite{GhoshIslamChemicalHardness2009}.

So far our discussion has been restricted to singlet-singlet excitations.
For the calculation of singlet-triplet excitations the only change required is in the coupling matrix elements, which for singlet-triplet excitations in TD-DFTB are given by
\begin{equation}\label{eq:CouplingMatrix_Triplet_TDDFTB}
K^\mathrm{T}_{ia,jb} = \sum_\mathcal{A} q_{ia,\mathcal A} \, W_\mathcal A \, q_{jb,\mathcal A} \; .
\end{equation}
Here the so called magnetic Hubbard parameters~$W_\mathcal A$ are defined as
\begin{equation}
W_\mathcal A = \frac{1}{2} \left( \frac{\partial \varepsilon^\mathrm{HOMO}_\spinup}{\partial n^\mathrm{HOMO}_\spinup} - \frac{\partial \varepsilon^\mathrm{HOMO}_\spinup}{\partial n^\mathrm{HOMO}_\spindown} \right) \; .
\end{equation}
and can be calculated from atomic DFT calculations just like the chemical hardnesses.

In addition to the approximation of the coupling matrix, TD-DFTB also approximates the transition dipole moments~$\vec d_{ia}$ of the single orbital transitions.
With the monopole approximation of the transition density from equation~\eqref{eq:MonopoleApproximation_TDDFTB} the transition dipole moments of the single orbital transitions are easily written as~\cite{NiehausTDDFTB2001}
\begin{equation}
\label{eq:SOTDM_TDDFTB} \vec d_{ia} = \sum_\mathcal A q_{ia,\mathcal A} \vec R_\mathcal A \; .
\end{equation}

One rather obvious limitation of the monopole approximation in equation~\eqref{eq:MonopoleApproximation_TDDFTB} is that basis functions~$\chi_\mu$ and~$\chi_\nu$ residing on the same atom~$\mathcal A$ do not contribute to the atomic transition charge~$q_{ia,\mathcal A}$.
This leads to vanishing (or underestimated) transition charges for excitations involving localized molecular orbitals~$\phi_i$ and~$\phi_a$, such as $\sigma \rightarrow \pi^*$ and $n \rightarrow \pi^*$ promotions.
Due to the vanishing coupling matrix elements~$K_{ia,jb}$ these excitations are then predicted to be pure single orbital transitions~$\phi_i \rightarrow \phi_a$ with an excitation energy~$\Delta_I = \Delta_{ia}$ exactly.
Furthermore, their transition dipole moment~$\vec d_I = \vec d_{ia}$ is predicted to be zero.
This failure has recently been corrected by \citeauthor{NiehausTDDFTBOnsiteAndFracOcc2013} through inclusion of one-center integrals of the exchange type~\cite{NiehausTDDFTBOnsiteAndFracOcc2013}.
However, this so-called on-site correction to TD-DFTB is fairly involved and we will restrict our discussion to TD-DFTB in its original formulation~\cite{NiehausTDDFTB2001}.

In summary, TD-DFTB is an approximation to TD-DFT that uses molecular orbitals obtained from a DFTB ground state calculation and approximates the coupling matrix and single orbital transition dipole moments in order to avoid integral evaluation at run-time.
For a recent review of TD-DFTB we would like to refer the reader to reference~\citenum{NiehausTDDFTBReview2009}.

\section{\label{s:TD-DFT+TB}TD-DFT+TB}

\subsection{Motivation and introduction}

In subsection~\ref{ss:TDDFTB} we have outlined how the TD-DFTB coupling matrix can be derived from its TD-DFT counterpart by making a monopole approximation for the transition density.
While this is how TD-DFTB was originally introduced~\cite{NiehausTDDFTB2001}, it is interesting to note that the same equations can also be obtained as the linear response of the SCC-DFTB Hamiltonian~\cite{NiehausTDDFTBOnsiteAndFracOcc2013}, just like TD-DFT was obtained from the linear response of DFT~\cite{CasidaTDDFT1995}.
In this sense all of the approximations that go into TD-DFTB have been done at the ground state level, and the subsequent excited state calculation is merely done consistently with the already present approximations.

\begin{figure}[tbp]
\includegraphics[width=0.95\columnwidth]{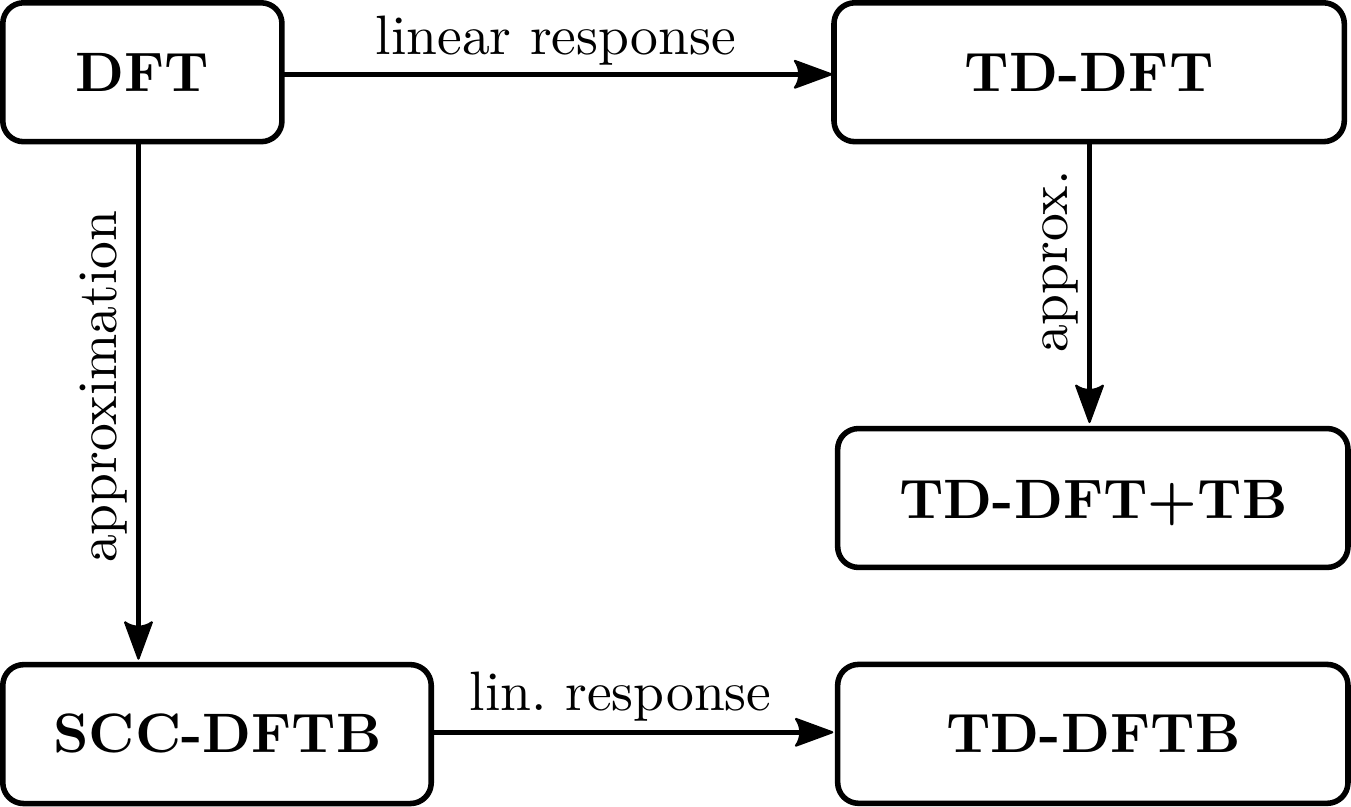}
\caption{\label{fig:method_comparison}Relationships among the different computational methods.}
\end{figure}

This brings up an interesting question:
Would more accurate results be obtained if the approximation was delayed until the linear response treatment?
Or in other words, would it be better to do an approximate linear response of the DFT Hamiltonian than to look at the exact linear response of the SCC-DFTB Hamiltonian?
In this article we want to propose to do TD-DFTB-like approximations in a linear response excited state calculation based on a DFT ground state.
We will henceforth refer to this approach as TD-DFT+TB.
The relationship between the different methods is illustrated in figure~\ref{fig:method_comparison}.

The basic idea of TD-DFT+TB is to use the molecular orbitals obtained from a DFT ground state calculation as input to an excited state calculation with the TD-DFTB coupling matrix from equation~\ref{eq:CouplingMatrix_TDDFTB}.
Technically, this is very easy to do: Looking back at subsection~\ref{ss:TDDFTB} it is evident that the only information needed about the ground state is the overlap matrix~$\mat S$, the coefficient matrix~$\mat C$ as well as the orbital energies~$\varepsilon_i$ and occupations.
Additionally, the information about which atom~$\mathcal A$ the basis function~$\chi_\mu$ is centered on is also needed for the population analysis.
However, all of this information could also be provided by a DFT instead of a DFTB ground state calculation.

One important thing to note is that the basis sets used in DFT are typically larger than the minimal basis set used in DFTB.
In fact, the pre-optimized DFTB ground state densities are typically of higher quality compared to those obtained using a minimum basis set DFT approach.
Therefore, it is important to employ a DFT basis that gives a sufficiently accurate ground state, even though this leads to more virtual orbitals and hence a larger coupling matrix in TD-DFT+TB compared to TD-DFTB.

A problem associated with the larger basis set in DFT is that the Mulliken population analysis~\cite{MullikenPopulationAnalysis1955} used in TD-DFTB for the calculation of the atomic transition charges~$q_{ia,\mathcal A}$ is known to become unstable for large basis sets, especially if diffuse basis functions are included.
While Mulliken analysis is working sufficiently well for the minimal atomic orbital basis set used in TD-DFTB, we have found that for a basis of TZP quality Mulliken transition charges only poorly represent the transition density.
This was also observed by~\citeauthor{GrimmeSimplifiedTDA2013}, who instead proposed~\cite{GrimmeSimplifiedTDA2013} to use L\"owdin population analysis~\cite{LowdinPopulationAnalysis1950} for which the atomic transition charges are calculated as
\begin{equation}\label{eq:AtomTransCharge_Loewdin}
q_{ia,\mathcal A} = \sum_{\mu \in \mathcal A} c'_{\mu i} c'_{\mu a} \quad \text{with} \quad \mat C' = \mat S^\frac{1}{2} \mat C
\; .
\end{equation}
We have indeed found L\"owdin transition charges to be much more reliable than the ones obtained from Mulliken population analysis, and therefore use L\"owdin analysis as the default method of calculating transition charges in TD-DFT+TB.
Benchmark results for both Mulliken and L\"owdin transition charges can be found in section~\ref{ss:vertexen}.

In TD-DFTB the atomic transition charges~$q_{ia,\mathcal A}$ are used in both the approximation of the $(N_\mathrm{occ} \times N_\mathrm{virt})^2$~coupling matrix elements as well as the approximation of the $(N_\mathrm{occ} \times N_\mathrm{virt})$~single orbital transition dipole moments~$\vec d_{ia}$.
While the former is what makes TD-DFTB so efficient, the latter has mostly technical reasons:
The Slater-Koster files used in DFTB contain the matrix elements, but not the basis functions themselves, making it impossible to evaluate integrals over molecular orbitals at run time.
This is a rather unpleasant deficiency introduced by the traditional DFTB Slater-Koster implementation.
However, it is not a deficiency of the method itself, and with knowledge of the atomic orbitals obtained during the parameterization process and in combination with a suitable integral engine any expectation value can be calculated correctly using DFTB.
This has been demonstrated for other properties, e.g.\ for NMR chemical shifts~\cite{HeineDFTBFullereneNMR1999}, and will be applied to the transition dipole matrix elements here.

For TD-DFT+TB, we have two possibilities to compute the transition dipole moments:
\begin{enumerate}
\item The simplest, and most approximate, way is to use the point charge approximation as done in TD-DFTB.
This approximation would be most attractive if TD-DFT+TB would be used by employing two independent codes (one for DFT and one for TD-DFTB).
However, in our present case there is negligible computational performance gain, so we do not follow this line.
\item We calculate the transition dipole moments directly from the DFT molecular orbitals.
Thus, we calculate the unapproximated single orbital transition dipole moments~$\vec d_{ia}$ from equation~\eqref{eq:SOTDM_exact}.
In different words: We avoid here those TD-DFTB approximations that have been due to restrictions imposed by the Slater-Koster-type implementation and which are not resulting in significant performance gain.
\end{enumerate}

One particularly attractive feature of TD-DFT+TB is that it does not rely at all on the DFTB parametrization.
The only parameters used for the construction of the TD-DFTB coupling matrix are the chemical hardness~$\eta_\mathcal{A}$ (for singlet excitations) and the magnetic Hubbard~$W_\mathcal{A}$ parameter (for triplet excitations).
These are just physical properties of the atoms that can be calculated and tabulated for the entire periodic table.
We use the chemical hardness as tabulated by \citeauthor{GhoshIslamChemicalHardness2009}~\cite{GhoshIslamChemicalHardness2009} and have calculated the values for the magnetic Hubbard parameter~$W_\mathcal{A}$ using the same details as specified earlier~\cite{WahiduzzamanQuasinano2013}.
The numerical values of $W_\mathcal{A}$ are given in table~\ref{tab:parameters}.
All other parameters entering DFTB which are needed for describing the ground state, i.e.\ the form of the basis functions, the effective potential, and the repulsive potential needed for calculating the total energy and its gradients are not needed to build the TD-DFTB coupling matrix.
TD-DFT+TB is therefore directly applicable to systems containing any combination of elements without the need of further parameterization.

\begin{table}[tbp]
\caption{\label{tab:method comparison}Comparison of the methods.}\vspace{8pt}
\resizebox{\columnwidth}{!}{\begin{tabular}{c|c|c|c}
 & TD-DFT & TD-DFT+TB & TD-DFTB \\ \hline\hline
 Molecular orbitals & \multicolumn{2}{c|}{from DFT} & from DFTB \\\hline
Coupling matrix~$\mat K$ & Eq.~\eqref{eq:CouplingMatrix_TDDFT} & \multicolumn{2}{c}{Eq.~\eqref{eq:CouplingMatrix_TDDFTB}} \\\hline
\specialcell{Atomic transition\\charges~$q_{ia,\mathcal A}$} & not used & Eq.~\eqref{eq:AtomTransCharge_Loewdin} & Eq.~\eqref{eq:AtomTransCharge_Mulliken} \\\hline
\specialcell{Single orbital transition\\dipole moments~$\vec d_{ia}$} & \multicolumn{2}{c|}{Eq.~\eqref{eq:SOTDM_exact}} & Eq.~\eqref{eq:SOTDM_TDDFTB} \\\hline
\specialcell{Chemical hardness~$\eta_\mathcal{A}$ \&\\Magnetic hubbard~$W_\mathcal{A}$} & not used & \multicolumn{2}{c}{\specialcell{precalculated by DFT\\for spherical atoms\footnote{included in the DFTB parameters files in case of TD-DFTB}}}
\end{tabular}}
\end{table}

In summary, TD-DFT+TB can be interpreted either as applying DFTB approximations to the \citeauthor{CasidaTDDFT1995} equations, or, equivalently, as TD-DFTB based on molecular orbitals from DFT.
Technical choices are the calculation of charges and transition dipole moments.
We propose to employ L\"owdin instead of Mulliken atomic transition charges, and DFT transition dipole moments, but other options are definitely possible.
A summarizing comparison of TD-DFT, TD-DFTB and TD-DFT+TB is given in table~\ref{tab:method comparison}.

\subsection{\label{ss:method_relation}Relation to other methods}

TD-DFT+TB as introduced in the last subsection is quite closely related to the sTDA~\cite{GrimmeSimplifiedTDA2013,GrimmeRangeSepsTDA2014} and sTD-DFT~\cite{GrimmeSimplifiedTDDFT2014} methods developed by \citeauthor{GrimmeSimplifiedTDA2013} and coworkers:
These methods also use molecular orbitals from a DFT calculation and use the same atomic monopole approximation for the transition density (which was originally introduced with TD-DFTB) in order to avoid the calculation of integrals.
The major difference is that TD-DFT+TB is a pure density functional approach, while sTDA and sTD-DFT use hybrid exchange-correlation functionals~\cite{BeckeHybrids1993} in both the calculation of the ground state and the excited states.

The primary reason why hybrid functionals with a fraction of exact Hartree-Fock exchange are often used in TD-DFT is that local functionals are known to drastically underestimate the excitation energies of charge-transfer excitations.
It was shown~\cite{GritsenkoCTFailure2004,BaerendsKSOrbitals2013,BaerendsKSOrbitals2014} that this failure can be traced back to the different meaning of virtual orbital energies in Kohn-Sham DFT and Hartree-Fock:
In DFT the virtual orbitals represent excited electrons interacting with $N-1$ other electrons, while in Hartree-Fock the virtual orbitals experience interaction with $N$~electrons, so that they represent added rather than excited electrons.
In other words, the Kohn-Sham HOMO-LUMO gap corresponds to the optical gap, while the Hartree-Fock HOMO-LUMO gap corresponds the fundamental gap, which is the difference between ionization energy and electron affinity.
It is easy to see why this leads to underestimated charge-transfer excitation energies in DFT:
If occupied and virtual orbital involved in a transition are localized on different fragments of the system, the transfered electron is essentially added to the acceptor fragment and its energy is determined by the acceptor's fundamental gap, not by its optical gap.
The fundamental gap is always larger than the optical gap, the difference being the interaction between the excited electron and the hole in its (now unoccupied) original orbital~\cite{BaerendsKSOrbitals2013,BredasMindTheGap2014}.
In summary, local excitations are well described in Kohn-Sham DFT with local functionals, while charge-transfer excitations profit from admixture of exact exchange.
The so-called range-separated hybrid functionals~\cite{IikuraRangeSep2001,HendersonRangeSep2008,BaerRangeSepReview2010}, where the amount of exact exchange increases with electron-electron distance, reflect this.

It is interesting to note though, that charge-transfer excitations typically have very small oscillator strengths.
Looking at equation~\eqref{eq:SOTDM_exact} it is easy to see that the transition dipole moment is zero if the involved orbitals~$\phi_i(\vec r)$ and~$\phi_a(\vec r)$  have no significant overlap, as is the case for charge-transfer excitations.
So even though charge-transfer excitation energies are severely underestimated in DFT with local functionals, the obtained absorption spectra are usually not affected.
There is, however, a technical problem associated with the underestimated charge transfer excitation energies for the specific application of calculating optical absorption spectra:
Since, the matrix~$\mat \Omega$ that has to be diagonalized in \citeauthor{CasidaTDDFT1995}'s equation~\eqref{eq:CasidasEquation} is extremely large, it is typically diagonalized using iterative eigensolvers that only calculate the few lowest eigenvectors.
If large numbers of spurious charge-transfer excitations are now predicted at much too low energies, many more excitations have to be calculated in order to cover the relevant energy range.
This drastically slows down the calculation even though the spurious charge-transfer excitations do not noticeably affect the obtained absorption spectra.
\citeauthor{GrimmeSimplifiedTDA2013} cites this issue as the main reason for the use of hybrid functionals in sTDA and sTD-DFT.
However, as we have recently shown the problem can also be solved by intensity selection~\cite{RugerIntensitySelection2015}, that is by simply neglecting single orbital transitions with small transition dipole moments.
This does not correct the energy of charge-transfer excitations but instead removes them from the spectrum altogether, leading to both a smaller number of excitations that have to be calculated as well as an overall smaller matrix~$\mat \Omega$ due to the reduced number of single orbital transitions.
While hybrid functionals are likely unavoidable if one needs accurate charge-transfer excitation energies, we believe that for the specific application of calculating absorption spectra, intensity selection is a much simpler and computationally more efficient alternative to hybrid functionals.

Furthermore, we would like to point out that while the use of hybrid functionals cures the charge-transfer problem, it introduces other problems that are not present in pure density functional approaches:
As pointed out by \citeauthor{BaerendsKSOrbitals2013} virtual orbitals from Kohn-Sham DFT with local functionals represent excited electrons interacting with $N-1$ other electrons.
The coupling matrix in equation~\eqref{eq:CasidaMatrixElements} is usually small compared to the orbital energy differences~$\Delta_{ia}$ on the diagonal, making orbital energy differences~$\Delta_{ia}$ an excellent approximation to excitation energies~$\Delta_I$~\cite{ZhangDFTOrbitalEigenvalues2007}.
Furthermore the excitations are often dominated by just one single orbital transition~$\phi_i \rightarrow \phi_a$, which drastically simplifies their interpretation~\cite{BaerendsKSOrbitals2013,BaerendsKSOrbitals2014}.
Hartree-Fock virtual orbitals on the other hand represent added electrons, so that their orbital energy differences~$\Delta_{ia}$ have little relation to excitation energies~$\Delta_I$ and are in fact much larger.
It is actually not uncommon for the Hartree-Fock LUMO to be unbound with an orbital energy of~$\varepsilon_a > 0$.
Furthermore, as the Hartree-Fock virtual orbitals interact with $N$ other electrons instead of $N-1$, they are much more diffuse than in DFT.
The Hartree-Fock virtuals are less suitable for the description of excited electrons and in general more of them are needed for the description of an excitation, meaning that excitations often lose the single orbital transition character they have in DFT, making their interpretation much more difficult~\cite{BaerendsKSOrbitals2013,BaerendsKSOrbitals2014}.
These problems are less severe if the employed exchange-correlation functional only has a small fraction of exact exchange.
It is, however, important to be aware of the fact that certain undesirable properties of time-dependent Hartree-Fock are reintroduced into TD-DFT if hybrid functionals are used.

In summary, we believe that there are good reasons to also approach excited state calculations for large systems from a pure density functional standpoint.

\section{\label{s:results}Method evaluation}

\subsection{\label{ss:vertexen}Vertical excitation energies}

In order to assess the accuracy of TD-DFT+TB we have calculated the lowest few excitation energies for the 28 molecules containing 1st and 2nd period elements in the benchmark set developed by \citeauthor{ThielTestset2008}~\cite{ThielTestset2008}.
For a direct comparison we have done the same calculations with TD-DFTB using the \texttt{mio-1-1} set of parameters~\cite{SeifertSCCDFTB1998,NiehausMIOSulfur2001,YangMIOPhosphorus2008,ElstnerDFTB32011}.
We use TD-DFT results as the reference against which TD-DFTB and TD-DFT+TB are compared.
Both TD-DFT and TD-DFT+TB results were obtained using the PBE exchange-correlation functional~\cite{PerdewBurkeErnzerhofPBEXCFunc1996} and a TZP~basis set.

Note that some excitations in the benchmark set by by \citeauthor{ThielTestset2008}~\cite{ThielTestset2008} have significant double excitation character and are hence difficult to describe with conventional TD-DFT. See reference~\citenum{Huix_Rotllant_2011} and~\citenum{Casida_2015} for a detailed discussion and possible solutions to this problem. However, for the purpose of comparing TD-DFT+TB and TD-DFTB to TD-DFT this does not play a role, as all methods are equally affected by this issue.

\begin{figure*}[p]
\includegraphics[height=\textwidth,angle=-90]{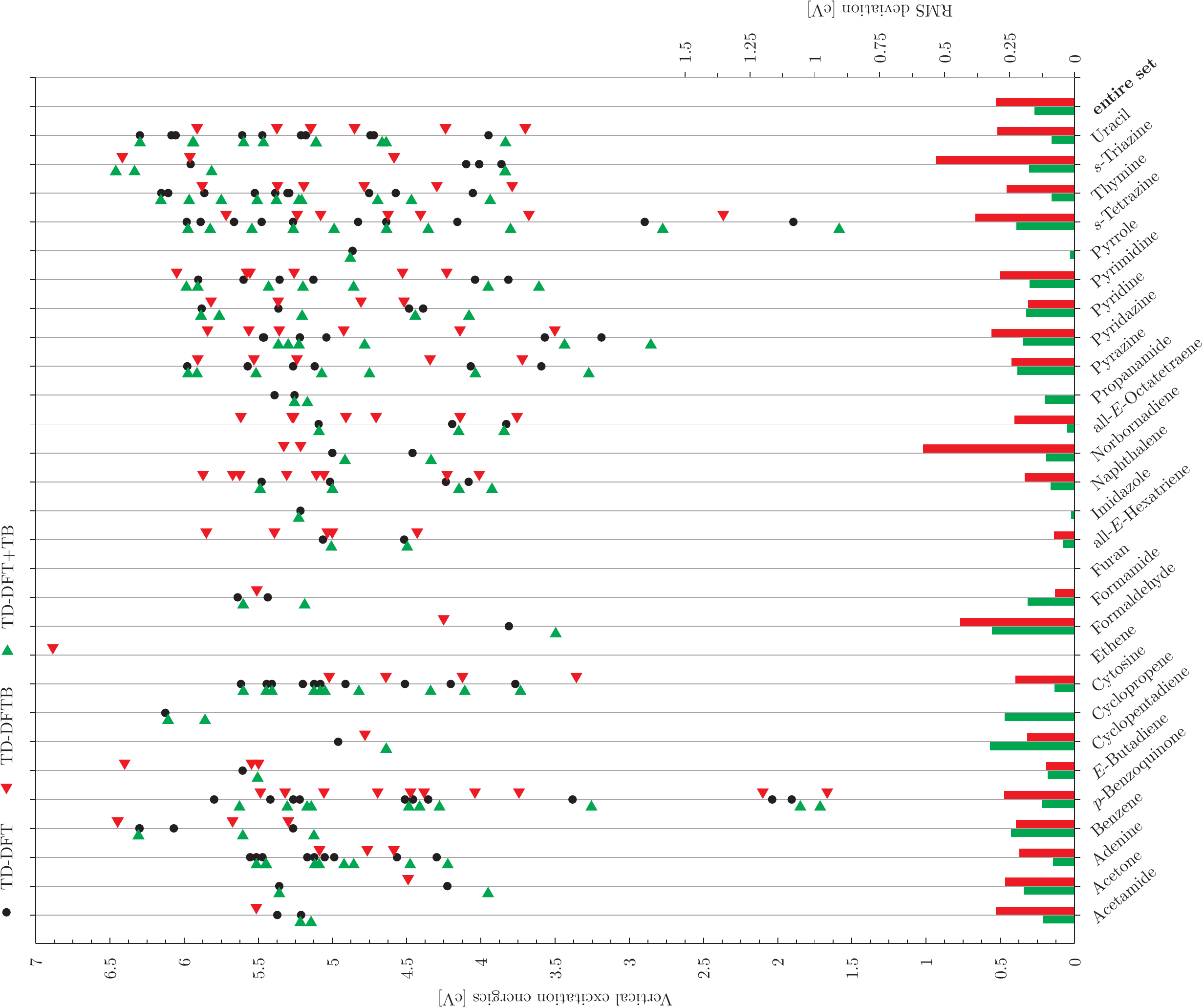}
\caption{\label{fig:thiel_singlet_loewdin}Vertical singlet-singlet excitation energies (left ordinate) for the molecules from \citeauthor{ThielTestset2008}'s test set~\cite{ThielTestset2008}. The bars at the bottom represent the RMSD in vertical excitation energies compared to TD-DFT (to scale with the right ordinate).}
\end{figure*}

\begin{figure*}[p]
\includegraphics[height=\textwidth,angle=-90]{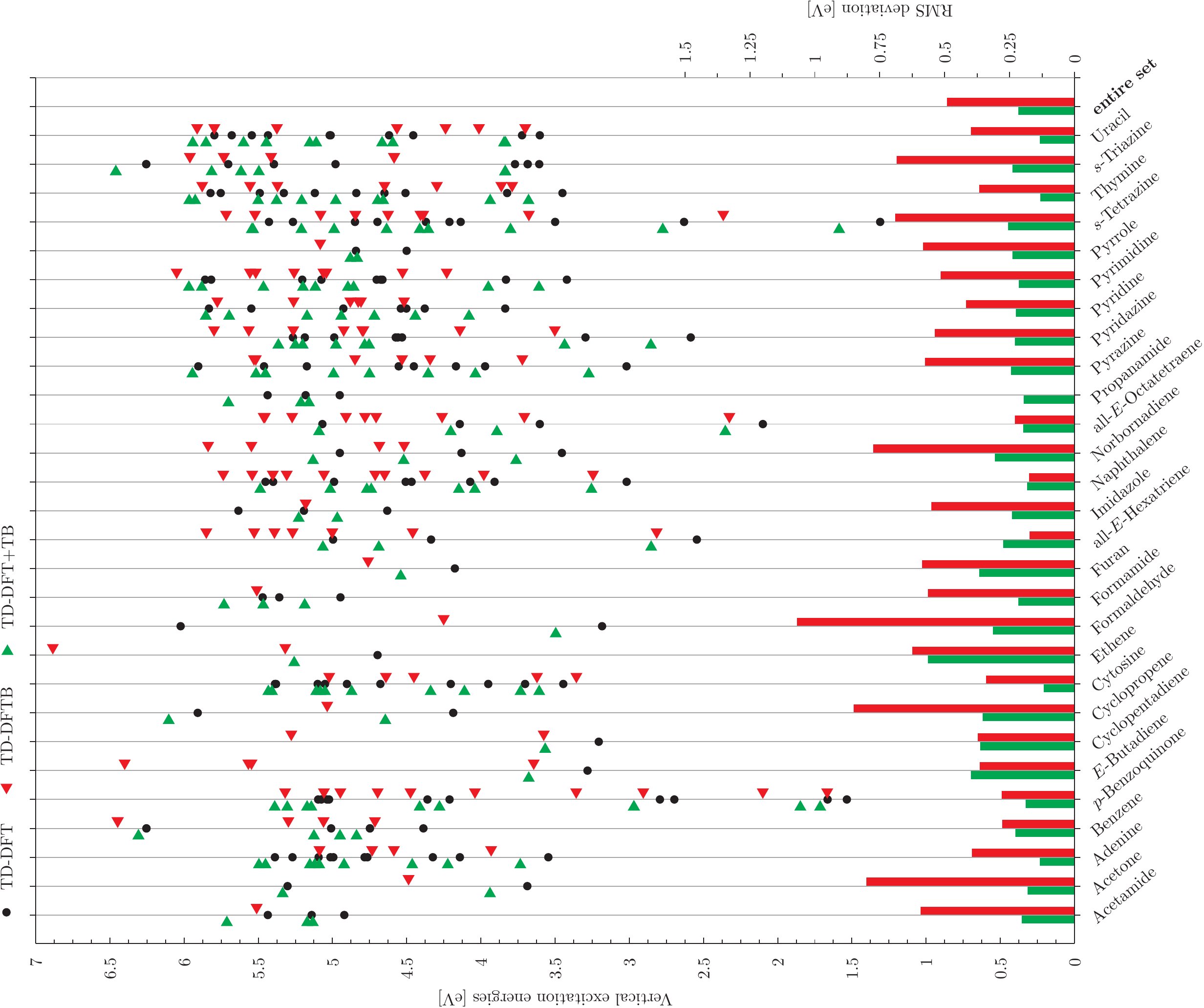}
\caption{\label{fig:thiel_triplet_loewdin}Vertical singlet-triplet excitation energies (left ordinate) for the molecules from \citeauthor{ThielTestset2008}'s test set~\cite{ThielTestset2008}. The bars at the bottom represent the RMSD in vertical excitation energies compared to TD-DFT (to scale with the right ordinate).}
\end{figure*}

The calculated vertical excitation energies and the root-mean-square deviation (RMSD) compared to TD-DFT are shown for the individual molecules in figure~\ref{fig:thiel_singlet_loewdin} and figure~\ref{fig:thiel_triplet_loewdin} for singlet-singlet and singlet-triplet excitations, respectively.
The calculated RMSD of all excitations in all molecules are shown in table~\ref{tab:thiel_statistics}.
Following \citeauthor{CasidaTDDFTLDALimits1998}'s recommendation~\cite{CasidaTDDFTLDALimits1998} we only considered excitations that have an excitation energy~$\Delta_I < -\varepsilon^\mathrm{HOMO}$ and no large contributions from transitions into unbound virtual orbitals with~$\varepsilon_a > 0$.
In cases where the number of excitations satisfying these criteria differs between the methods, we compare the lowest common number of excitations.
For singlet-singlet excitations in ethene and furan none of the calculated excitations satisfies both of \citeauthor{CasidaTDDFTLDALimits1998}'s criteria, so that these two molecules had to be excluded from the calculation of the overall RMSD for singlet-singlet excitations.

Compared to normal TD-DFTB, TD-DFT+TB is closer to TD-DFT for both singlet-singlet and singlet-triplet excitations.
For singlet-singlet excitations switching from TD-DFTB to TD-DFT+TB reduces the RMSD by a factor of two from 0.301eV to 0.153eV.
It is known that TD-DFTB is more accurate for singlet-singlet excitations than for singlet-triplet excitations~\cite{NiehausTDDFTB2001} for which we calculated an RMSD of 0.489eV.
We observe the same behavior for TD-DFT+TB, although with an RMSD of 0.215eV the difference in accuracy between singlet-singlet and singlet-triplet excitations is slightly smaller.

\begin{table}[tb]
\begin{tabular}{c|c|c}
Multiplicity & TD-DFTB & TD-DFT+TB \\\hline\hline
singlet-singlet & 0.30eV & 0.15eV \\ \hline
singlet-triplet & 0.49eV & 0.22eV
\end{tabular}
\caption{\label{tab:thiel_statistics}Root-mean-square deviation in vertical excitation energies of TD-DFTB and TD-DFT+TB for \citeauthor{ThielTestset2008}'s test set. TD-DFT is used as the reference method.}
\end{table}

\begin{table*}[tbp]
\begin{tabular}{c|c|c|c|c|c|c|c|c|c|c|c|c}
\multirow{2}{*}{transition} & \multicolumn{4}{c}{TD-DFT} & \multicolumn{4}{|c|}{TD-DFT+TB} & \multicolumn{4}{c}{TD-DFTB} \\
 & $\Delta_I$ & $f_I$ & $\Delta_{ia}$ & $\Delta_I \hspace{-3pt} - \hspace{-4pt} \Delta_{ia}$ & $\Delta_I$ & $f_I$ & $\Delta_{ia}$ & $\Delta_I \hspace{-3pt} - \hspace{-4pt} \Delta_{ia}$ & $\Delta_I$ & $f_I$ & $\Delta_{ia}$ & $\Delta_I \hspace{-3pt} - \hspace{-4pt} \Delta_{ia}$ \\ \hline \hline
$^1\Pi_u$ ($\sigma_u \rightarrow \pi_g$) & 13.47 & 0.32 & 11.53 & 1.94 & 11.53 & 0.64 & 11.53 & 0.00 & 12.73 & 0.00 & 12.73 & 0.00 \\ \hline
$^1\Sigma^+_u$ $(\pi_u \rightarrow \pi_g)$ & 14.94 & 0.34 & 9.55 & 4.94 & 12.99 & 0.98 & 9.55 & 3.44 & 13.90 & 0.88 & 10.19 & 3.71
\end{tabular}
\caption{\label{tab:N2_excitations}Dipole allowed transitions in N$_2$. All energies in eV. $\Delta_{ia}$ is the orbital energy difference for the dominant single orbital transition.}
\end{table*}

Note that for the calculation of the RMSD we have simply compared the excitation energies from the different methods according to their order in energy.
We have not attempted to compensate for the fact that two excited states might switch in energy ordering when going from TD-DFT to one of the approximate methods.
While this does not affect the comparison between TD-DFTB and TD-DFT+TB, the absolute errors in table~\ref{tab:thiel_statistics} will be slightly underestimated and one should be careful when comparing them to the literature.

As mentioned in section~\ref{s:TD-DFT+TB} we have also run TD-DFT+TB calculations for \citeauthor{ThielTestset2008}'s test set using Mulliken instead of L\"owdin population analysis for the calculation of the atomic transition charges.
We found an RMSD of 0.449eV in vertical singlet-singlet excitation energies, which is three times larger than the 0.153eV obtained with L\"owdin charges, indicating that Mulliken transition charges do not accurately model the transition density for the relatively large TZP basis set used.
For singlet-triplet excitations we have furthermore found that unphysical transition charges sometimes lead to negative eigenvalues in equation~\eqref{eq:CasidasEquation} and hence imaginary excitation energies.

\subsection{Oscillator strengths and absorption spectra}

In the last subsection we looked exclusively at vertical excitation energies.
However, for the application of calculating UV/Vis absorption spectra both excitation energies and oscillator strengths have to be calculated.

\subsubsection{N$_2$}

One difference between TD-DFTB and TD-DFT+TB is that the latter does not use the atomic transition charges for the calculation of the single orbital transition dipole moments.
To illustrate the effect of this we have calculated the lowest excitations in N$_2$ with a nuclear distance of 1.106\AA.
The results are shown in table~\ref{tab:N2_excitations}.
According to TD-DFT there are two dipole allowed transitions:
A $^1\Pi_u$~state consisting mainly of a $\sigma_u \rightarrow \pi_g$ transition, and a $^1\Sigma^+_u$~state dominated by a $\pi_u \rightarrow \pi_g$ transition.
Note that even though both of them have excitation energies $\Delta_I > -\varepsilon^\mathrm{HOMO}$ we have found them to be well described and largely basis set independent due to the fact that they do not have contributions from transitions into unbound virtual orbitals.
The $^1\Sigma^+_u$~state is reasonably well described by both TD-DFTB and TD-DFT+TB, who both predict it to be dipole-allowed.
Both methods underestimate the vertical excitation energy~$\Delta_I$ of the $^1\Sigma^+_u$~state, with the TD-DFTB energy being closer to the TD-DFT reference.
However, this is mostly due to the larger orbital energy difference~$\Delta_{ia}$ in DFTB compared to DFT, since the coupling matrix induced shift $\Delta_I - \Delta_{ia}$ is similar for both TD-DFTB and TD-DFT+TB.
The $\sigma_u \rightarrow \pi_g$ transition into the $^1\Pi_u$~state is less well described with the approximate methods.
TD-DFT predicts a coupling matrix induced shift~$\Delta_I - \Delta_{ia}$ of almost 2eV while both approximate methods produce exactly a single orbital transition with~$\Delta_I = \Delta_{ia}$.
This is due to the atomic transition charges' inability to model local transitions as mentioned in subsection~\ref{ss:TDDFTB}.
Since TD-DFTB also uses the transition charges for the transition dipole moments, it incorrectly predicts the transition into the $^1\Pi_u$~state state to be dipole-forbidden.
This is not the case in TD-DFT+TB, so that the method can at least be used to identify dipole-allowed $\sigma \rightarrow \pi^*$ and $n \rightarrow \pi^*$ transitions, even though their excitation energies will be less accurate than those of $\pi \rightarrow \pi^*$~transitions.

However, for the large systems such approximate method are typically used for, $\pi \rightarrow \pi^*$~transitions usually have the largest oscillator strengths, so that TD-DFTB and TD-DFT+TB's problems with localized transitions often do not noticeably affect the calculated absorption spectra.

\subsubsection{Fullerene C$_{60}$}

As an example for the calculation of absorption spectra, we have calculated the UV/Vis spectrum of the C$_{60}$~fullerene.
This was one of the example systems in the original TD-DFTB article and also makes a good technical benchmark as almost a thousand excitations have to be calculated to cover the relevant energy range.
For TD-DFT and TD-DFT+TB we used a TZP~basis and the PBE~exchange-correlation functional~\cite{PerdewBurkeErnzerhofPBEXCFunc1996}.
TD-DFTB calculations were performed with both the \texttt{3ob-3-1}~parameter set~\cite{Elstner3obParameters2013,Elstner3obSPParameters2014,Elstner3obMgZnParameters2015,Elstner3obMoreParameters2015} and the \texttt{QUASINANO2013} set by \citeauthor{WahiduzzamanQuasinano2013}~\cite{WahiduzzamanQuasinano2013}.
For calculations with the \texttt{3ob-3-1}~parameter set the ground state calculation was performed at the DFTB3~\cite{ElstnerDFTB32011} level of theory.
Conceptually this is slightly inconsistent, as the calculation of the excited states is based on the linear response of a Hamiltonian different from the one used for the calculation of the ground state.
However, since the DFTB3 orbitals are generally of better quality than DFTB2 orbitals, this gives rather good results in practice.

\begin{figure}[tbp]
\includegraphics[width=\columnwidth]{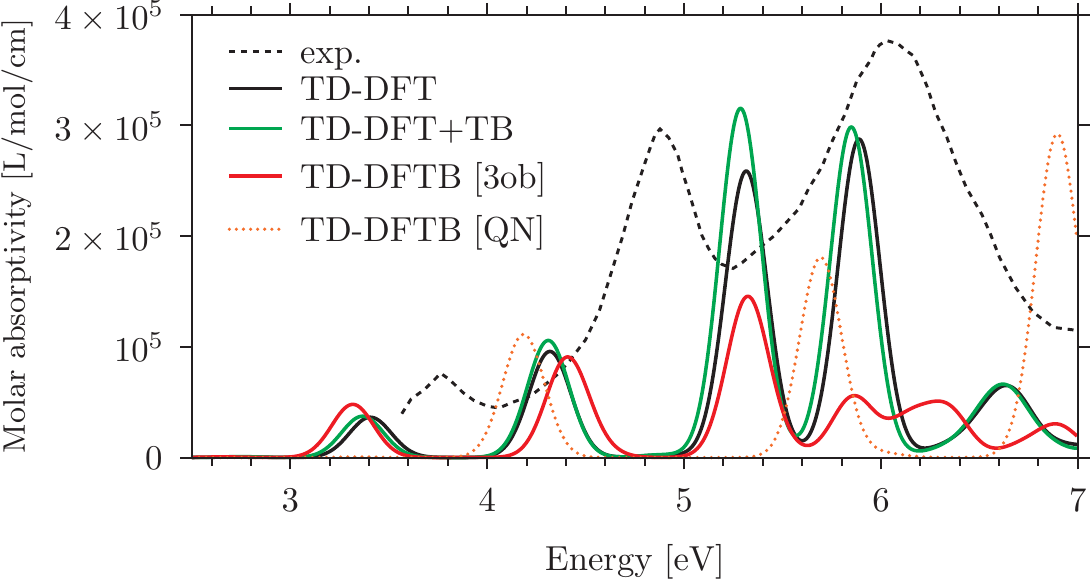}
\caption{\label{fig:fullerene_spectrum}Absorption spectrum of the C$_{60}$~fullerene. Experimental gas phase absorption spectrum from reference~\citenum{YasumatsuC60Abs1996}. Note that the authors quote a 100\% uncertainty in the absolute absorption cross sections due to the vapor pressure--temperature relation. Theoretical spectra have been broadened with a $\sigma = 0.25\mathrm{eV}$ Gaussian.}
\end{figure}

The calculated spectra are shown in figure~\ref{fig:fullerene_spectrum}.
TD-DFT+TB reproduces the TD-DFT reference spectrum almost perfectly.
TD-DFTB with the \texttt{3ob-3-1}~parameter performs very well below 5.5eV but underestimates the intensity of an excitation seen at 5.9eV in the TD-DFT spectrum.
All three spectra qualitatively reproduce the series of absorption bands of increasing intensity seen in the experimental spectrum~\cite{YasumatsuC60Abs1996}.
However, the theoretical spectra are redshifted compared to experiment.
Absolute intensities should not be compared to experiment, as the experimentally measured cross sections have an uncertainty of $100\%$ due to the sensitive vapor pressure--temperature relation of fullerenes~\cite{AbrefahC60VaporPressure1992}.

The TD-DFTB spectrum calculated with the \texttt{QUASINANO2013.1} parameters shows a substantial blue-shift compared to the other methods and the experimental reference.
However, the shape of the spectrum with its three bands of increasing oscillator strength is reasonably well described.
The origin of the blue-shift can be traced back to differences in the Kohn-Sham orbital energies:
DFT and DFTB with the \texttt{3ob-3-1}~parameters show a HOMO-LUMO gap of about 1.6eV, while DFTB with the \texttt{QUASINANO2013.1} parameters produces a gap of 2.3eV.
Keeping in mind that the HOMO-LUMO gap in DFT represents the optical gap~\cite{ZhangDFTOrbitalEigenvalues2007,BaerendsKSOrbitals2013,BaerendsKSOrbitals2014}, it is easy to understand why the \texttt{QUASINANO2013.1} parameters predict overall larger excitation energies.
The reason for the larger orbital energy differences with the \texttt{QUASINANO2013.1} parameters is that they were optimized to reproduce band structures in solids~\cite{WahiduzzamanQuasinano2013}, for which relatively tight confinement potentials are required in the atomic calculations.
However, the additional potential leads to increased orbital energy differences through quantum confinement and produces systematically overestimated excitation energies and blue-shifted absorption spectra.
This illustrates how strongly TD-DFTB results can depend on details of the DFTB parametrization; a problem that does not exist in TD-DFT+TB.

Timings for the calculation of the C$_{60}$~absorption spectra are shown in table~\ref{tab:fullerene_timings}.
\begin{table}[tb]
\begin{tabular}{c|c|c|c}
 & TD-DFT & TD-DFT+TB & TD-DFTB \\\hline\hline
ground state & 4min 38s & 4min 33s & < 1s \\\hline
excited states & 19h 37min & 11min 35s & 1min 26s
\end{tabular}
\caption{\label{tab:fullerene_timings}Timings for the calculation of the 988 lowest singlet-singlet excitations in the C$_{60}$~fullerene. The obtained spectra are shown in figure~\ref{fig:fullerene_spectrum}. All calculations were performed on an Intel Core i7-4770 processor.}
\end{table}
The benchmark TD-DFT calculation take almost 20~hours on a recent workstation computer, only 5~minutes of which are spent calculating the ground state.
With TD-DFT+TB the total wall-time decreases to about 16~minutes, which is a speedup by a factor of~73 compared to TD-DFT.
With a total wall-time of less than 90~seconds TD-DFTB still much faster that TD-DFT+TB.
The DFTB ground state calculation takes less than a second and is therefore completely negligible compared to the 5~minutes for the DFT ground state in TD-DFT+TB.
Furthermore, due to the minimal basis set the space of single orbital transitions is much smaller in TD-DFTB, leading to a smaller matrix to be diagonalized:
For DFT with a TZP basis set there are $120 \times 900 = 108000$ single orbital transitions, whereas DFTB only has $120 \times 120 = 14400$ transitions.

\subsubsection{Chlorophyll A}

\begin{figure}[tbp]
\includegraphics[width=\columnwidth]{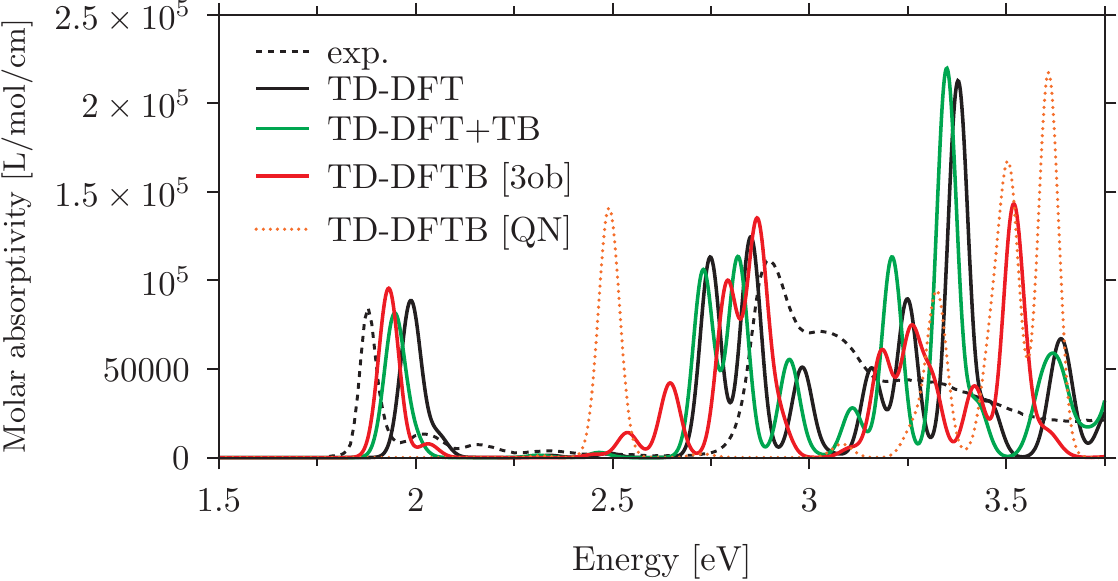}
\caption{\label{fig:chlA_spectrum}Absorption spectrum of chlorophyll~A. Experimental spectrum measured in diethyl ether prepared by Scott Prahl based on reference~\citenum{StrainChlA1963} and~\citenum{DuPhotochemCAD1998}. Theoretical spectra have been broadened with a $\sigma = 0.06\mathrm{eV}$ Gaussian.}
\end{figure}

We have also calculated the UV/Vis absorption spectrum of chlorophyll~A. Due to the magnesium ion at the center of the chlorin ring, DFTB parameters that allow calculations of chlorophyll have only recently become available~\cite{WahiduzzamanQuasinano2013,Elstner3obMgZnParameters2015}.
The calculated spectra are shown in figure~\ref{fig:chlA_spectrum}.
The agreement between TD-DFT and TD-DFT+TB is again almost perfect throughout the entire energy range and both methods show the well-known $Q_y$ and Soret absorption bands around 1.95eV and 2.8eV, respectively.
The spectrum obtained with TD-DFTB and the \texttt{3ob-3-1}~parameters is very close to TD-DFT below 3.2eV, but differs somewhat beyond that.
All three methods reproduce the essential features of the experimental absorption spectrum~\cite{StrainChlA1963,DuPhotochemCAD1998}, although the energy gap between $Q_y$ and Soret band is slightly underestimated.
Note, however, that the experimental spectrum was recorded in solution, while our calculation corresponds to absorption in the gas phase.
In the region above $3\mathrm{eV}$ the theoretical spectra show more structure than the relatively flat experimental spectrum, which we attribute to the neglect of vibrational broadening in the theoretical spectra.
With the \texttt{QUASINANO2013} parameters we again observe a blue-shift of the entire spectrum, so that $Q_y$ and Soret band are predicted at 2.5eV and 3.6eV respectively.

\subsubsection{Ir(ppy)$_3$}

\begin{figure}[tbp]
\includegraphics[width=\columnwidth]{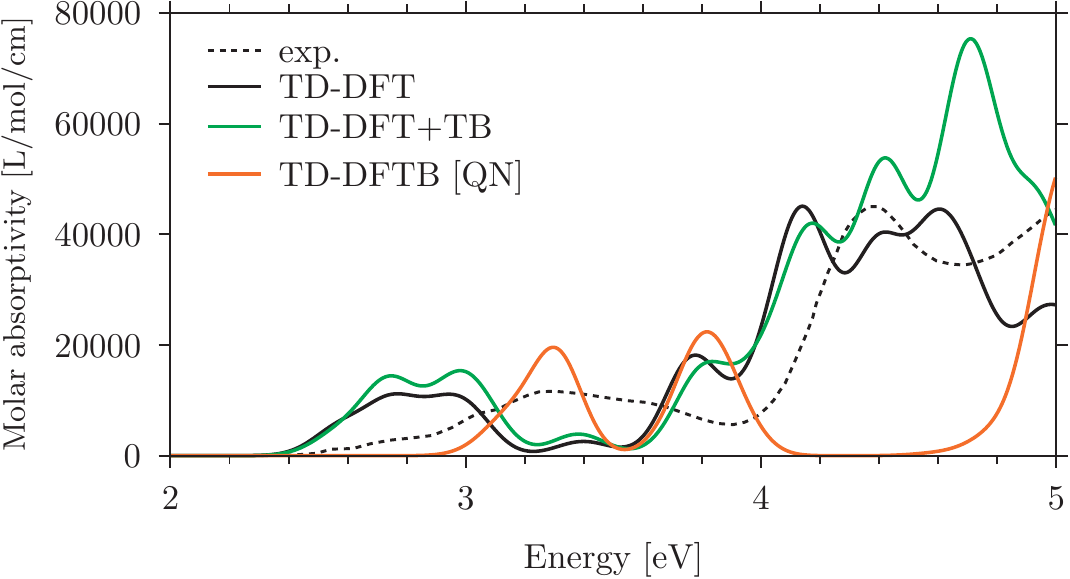}
\caption{\label{fig:irppy3_spectrum}Absorption spectrum of Ir(ppy)$_3$. Experimental spectrum measured in dichloromethane from reference~\citenum{FineIrppy3Spectrum2012}. Note that the experimental reference does not give absolute absorptivities. The experimental spectrum was therefore scaled to reproduce the TD-DFT value at the peak just above~$4\mathrm{eV}$. Theoretical spectra have been broadened with a $\sigma = 0.2\mathrm{eV}$ Gaussian.}
\end{figure}

Our last example calculation is the UV/Vis absorption spectrum of \textit{fac}-Ir(ppy)$_3$ (an abbreviation for \textit{fac}-Tris(2-phenylpyridine)iridium), a compound that is of interest in the context of highly efficient organic light emitting diodes~\cite{irppy3_0}.
The calculated absorption spectra are shown in figure~\ref{fig:irppy3_spectrum}.
Below 4.4eV TD-DFT+TB agrees well with the TD-DFT reference spectrum.
Beyond that energy range the oscillator strengths from TD-DFT+TB seem to be overestimated, so that the predicted absorption is overall too strong.
Both methods reproduce the principal features of the experimentally measured spectrum~\cite{FineIrppy3Spectrum2012}, though the absorption spectra are slightly redshifted compared to experiment.
Note that the experimental spectrum was measured in solution, while our calculations correspond to gas phase absorption.
Absolute experimental absorption coefficients were not given in reference~\citenum{FineIrppy3Spectrum2012} and can hence not be compared to theory.
Due to the iridium atom at the center of the complex, DFTB calculations of Ir(ppy)$_3$ can at the moment only be performed with the \texttt{QUASINANO2013} set of parameters.
For both the fullerene and the chlorophyll we had observed a blue-shift in the spectra calculated with these parameters, which is again the case for Ir(ppy)$_3$.

\section{\label{s:conclusion}Conclusion}

In summary we have presented a new method for calculating electronic excitations that combines molecular orbitals from a DFT ground state calculation with TD-DFTB like approximations for the coupling matrix from \citeauthor{CasidaTDDFT1995}'s linear response formalism.
We have shown that the new method named TD-DFT+TB improves vertical excitations energies compared to TD-DFTB and yields electronic absorption spectra that almost perfectly agree with computationally much more costly TD-DFT calculations.
In contrast to TD-DFTB, TD-DFT+TB does not rely on DFTB parametrization and is therefore applicable to molecular systems containing any combination of elements.

The new method is very easy to implement into existing DFT codes that already have support for TD-DFT, since it is essentially only a simplification of the coupling matrix.
Alternatively it could also very easily be supported by a standalone DFTB implementation with TD-DFTB support:
Instead of calculating the molecular orbitals using DFTB, one could read orbitals calculated by an external DFT code from disk and use them as input for the TD-DFTB calculation.
While both approaches are viable, we believe that direct integration into a DFT code is the more user-friendly alternative.
We have integrated TD-DFT+TB in this way into the 2016 release of the ADF modeling suite~\cite{ADF2001}.

Our method is implemented and can be used for a wide range of systems where TD-DFT is computationally unfeasible.
However, there is still room for improvements, and we are currently working on several enhancements and further validations.
For heavier elements we are currently investigating whether orbital-dependent hardness parameters give superior performance compared to the presently used atomic ones.
We are further assessing the performance of the approach by comparing smaller (DZP) and larger (TZ2P) basis sets.

Looking at the bigger picture, there are by now several related methods for the calculation of exited state properties of large systems, namely TD-DFTB, TD-DFT+TB, sTDA and sTD-DFT.
It would be desirable to benchmark all these different methods against experimental data in order to be able to give clear recommendations to end users, regarding applicability and accuracy of the various methods.
Based on the discussion in section~\ref{ss:method_relation}, we for example expect intensity selected TD-DFT+TB to be very suitable for the calculation and analysis of absorption spectra, while sTD-DFT should yield generally more accurate excitation energies.
We believe that a consistently done benchmark study including all the different methods could serve to give end users the right tools for the right applications and would in general make such methods more approachable for non-expert users.

\begin{acknowledgements}
The research leading to these results has received funding from the European Union's Seventh Framework Programme (FP7-PEOPLE-2012-ITN) under project PROPAGATE, GA~316897.
\end{acknowledgements}

\bibliography{literature,paper1,paper3}

\begin{thebibliography}{69}%
\makeatletter
\providecommand \@ifxundefined [1]{%
 \@ifx{#1\undefined}
}%
\providecommand \@ifnum [1]{%
 \ifnum #1\expandafter \@firstoftwo
 \else \expandafter \@secondoftwo
 \fi
}%
\providecommand \@ifx [1]{%
 \ifx #1\expandafter \@firstoftwo
 \else \expandafter \@secondoftwo
 \fi
}%
\providecommand \natexlab [1]{#1}%
\providecommand \enquote  [1]{``#1''}%
\providecommand \bibnamefont  [1]{#1}%
\providecommand \bibfnamefont [1]{#1}%
\providecommand \citenamefont [1]{#1}%
\providecommand \href@noop [0]{\@secondoftwo}%
\providecommand \href [0]{\begingroup \@sanitize@url \@href}%
\providecommand \@href[1]{\@@startlink{#1}\@@href}%
\providecommand \@@href[1]{\endgroup#1\@@endlink}%
\providecommand \@sanitize@url [0]{\catcode `\\12\catcode `\$12\catcode
  `\&12\catcode `\#12\catcode `\^12\catcode `\_12\catcode `\%12\relax}%
\providecommand \@@startlink[1]{}%
\providecommand \@@endlink[0]{}%
\providecommand \url  [0]{\begingroup\@sanitize@url \@url }%
\providecommand \@url [1]{\endgroup\@href {#1}{\urlprefix }}%
\providecommand \urlprefix  [0]{URL }%
\providecommand \Eprint [0]{\href }%
\providecommand \doibase [0]{http://dx.doi.org/}%
\providecommand \selectlanguage [0]{\@gobble}%
\providecommand \bibinfo  [0]{\@secondoftwo}%
\providecommand \bibfield  [0]{\@secondoftwo}%
\providecommand \translation [1]{[#1]}%
\providecommand \BibitemOpen [0]{}%
\providecommand \bibitemStop [0]{}%
\providecommand \bibitemNoStop [0]{.\EOS\space}%
\providecommand \EOS [0]{\spacefactor3000\relax}%
\providecommand \BibitemShut  [1]{\csname bibitem#1\endcsname}%
\let\auto@bib@innerbib\@empty
\bibitem [{\citenamefont {Hohenberg}\ and\ \citenamefont
  {Kohn}(1964)}]{HohenbergKohnTheorem1964}%
  \BibitemOpen
  \bibfield  {author} {\bibinfo {author} {\bibfnamefont {P.}~\bibnamefont
  {Hohenberg}}\ and\ \bibinfo {author} {\bibfnamefont {W.}~\bibnamefont
  {Kohn}},\ }\href {\doibase 10.1103/PhysRev.136.B864} {\bibfield  {journal}
  {\bibinfo  {journal} {Phys. Rev.}\ }\textbf {\bibinfo {volume} {136}},\
  \bibinfo {pages} {B864} (\bibinfo {year} {1964})}\BibitemShut {NoStop}%
\bibitem [{\citenamefont {Kohn}\ and\ \citenamefont
  {Sham}(1965)}]{KohnShamEquations1965}%
  \BibitemOpen
  \bibfield  {author} {\bibinfo {author} {\bibfnamefont {W.}~\bibnamefont
  {Kohn}}\ and\ \bibinfo {author} {\bibfnamefont {L.~J.}\ \bibnamefont
  {Sham}},\ }\href {\doibase 10.1103/PhysRev.140.A1133} {\bibfield  {journal}
  {\bibinfo  {journal} {Phys. Rev.}\ }\textbf {\bibinfo {volume} {140}},\
  \bibinfo {pages} {A1133} (\bibinfo {year} {1965})}\BibitemShut {NoStop}%
\bibitem [{\citenamefont {Porezag}\ \emph {et~al.}(1995)\citenamefont
  {Porezag}, \citenamefont {Frauenheim}, \citenamefont {K\"ohler},
  \citenamefont {Seifert},\ and\ \citenamefont {Kaschner}}]{PorezagDFTB1995}%
  \BibitemOpen
  \bibfield  {author} {\bibinfo {author} {\bibfnamefont {D.}~\bibnamefont
  {Porezag}}, \bibinfo {author} {\bibfnamefont {T.}~\bibnamefont {Frauenheim}},
  \bibinfo {author} {\bibfnamefont {T.}~\bibnamefont {K\"ohler}}, \bibinfo
  {author} {\bibfnamefont {G.}~\bibnamefont {Seifert}}, \ and\ \bibinfo
  {author} {\bibfnamefont {R.}~\bibnamefont {Kaschner}},\ }\href {\doibase
  10.1103/PhysRevB.51.12947} {\bibfield  {journal} {\bibinfo  {journal} {Phys.
  Rev. B}\ }\textbf {\bibinfo {volume} {51}},\ \bibinfo {pages} {12947}
  (\bibinfo {year} {1995})}\BibitemShut {NoStop}%
\bibitem [{\citenamefont {Seifert}, \citenamefont {Porezag},\ and\
  \citenamefont {Frauenheim}(1996)}]{SeifertDFTB1996}%
  \BibitemOpen
  \bibfield  {author} {\bibinfo {author} {\bibfnamefont {G.}~\bibnamefont
  {Seifert}}, \bibinfo {author} {\bibfnamefont {D.}~\bibnamefont {Porezag}}, \
  and\ \bibinfo {author} {\bibfnamefont {T.}~\bibnamefont {Frauenheim}},\
  }\href {\doibase 10.1002/(SICI)1097-461X(1996)58:2<185::AID-QUA7>3.0.CO;2-U}
  {\bibfield  {journal} {\bibinfo  {journal} {Int. J. Quantum Chem.}\ }\textbf
  {\bibinfo {volume} {58}},\ \bibinfo {pages} {185} (\bibinfo {year}
  {1996})}\BibitemShut {NoStop}%
\bibitem [{\citenamefont {Slater}\ and\ \citenamefont
  {Koster}(1954)}]{SlaterKoster1954}%
  \BibitemOpen
  \bibfield  {author} {\bibinfo {author} {\bibfnamefont {J.~C.}\ \bibnamefont
  {Slater}}\ and\ \bibinfo {author} {\bibfnamefont {G.~F.}\ \bibnamefont
  {Koster}},\ }\href {\doibase 10.1103/PhysRev.94.1498} {\bibfield  {journal}
  {\bibinfo  {journal} {Phys. Rev.}\ }\textbf {\bibinfo {volume} {94}},\
  \bibinfo {pages} {1498} (\bibinfo {year} {1954})}\BibitemShut {NoStop}%
\bibitem [{\citenamefont {Elstner}\ \emph {et~al.}(1998)\citenamefont
  {Elstner}, \citenamefont {Porezag}, \citenamefont {Jungnickel}, \citenamefont
  {Elsner}, \citenamefont {Haugk}, \citenamefont {Frauenheim}, \citenamefont
  {Suhai},\ and\ \citenamefont {Seifert}}]{SeifertSCCDFTB1998}%
  \BibitemOpen
  \bibfield  {author} {\bibinfo {author} {\bibfnamefont {M.}~\bibnamefont
  {Elstner}}, \bibinfo {author} {\bibfnamefont {D.}~\bibnamefont {Porezag}},
  \bibinfo {author} {\bibfnamefont {G.}~\bibnamefont {Jungnickel}}, \bibinfo
  {author} {\bibfnamefont {J.}~\bibnamefont {Elsner}}, \bibinfo {author}
  {\bibfnamefont {M.}~\bibnamefont {Haugk}}, \bibinfo {author} {\bibfnamefont
  {T.}~\bibnamefont {Frauenheim}}, \bibinfo {author} {\bibfnamefont
  {S.}~\bibnamefont {Suhai}}, \ and\ \bibinfo {author} {\bibfnamefont
  {G.}~\bibnamefont {Seifert}},\ }\href {\doibase 10.1103/PhysRevB.58.7260}
  {\bibfield  {journal} {\bibinfo  {journal} {Phys. Rev. B}\ }\textbf {\bibinfo
  {volume} {58}},\ \bibinfo {pages} {7260} (\bibinfo {year}
  {1998})}\BibitemShut {NoStop}%
\bibitem [{\citenamefont {Gaus}, \citenamefont {Cui},\ and\ \citenamefont
  {Elstner}(2011)}]{ElstnerDFTB32011}%
  \BibitemOpen
  \bibfield  {author} {\bibinfo {author} {\bibfnamefont {M.}~\bibnamefont
  {Gaus}}, \bibinfo {author} {\bibfnamefont {Q.}~\bibnamefont {Cui}}, \ and\
  \bibinfo {author} {\bibfnamefont {M.}~\bibnamefont {Elstner}},\ }\href
  {\doibase 10.1021/ct100684s} {\bibfield  {journal} {\bibinfo  {journal} {J.
  Chem. Theory Comput.}\ }\textbf {\bibinfo {volume} {7}},\ \bibinfo {pages}
  {931} (\bibinfo {year} {2011})}\BibitemShut {NoStop}%
\bibitem [{\citenamefont {Gaus}, \citenamefont {Goez},\ and\ \citenamefont
  {Elstner}(2013)}]{Elstner3obParameters2013}%
  \BibitemOpen
  \bibfield  {author} {\bibinfo {author} {\bibfnamefont {M.}~\bibnamefont
  {Gaus}}, \bibinfo {author} {\bibfnamefont {A.}~\bibnamefont {Goez}}, \ and\
  \bibinfo {author} {\bibfnamefont {M.}~\bibnamefont {Elstner}},\ }\href
  {\doibase 10.1021/ct300849w} {\bibfield  {journal} {\bibinfo  {journal} {J.
  Chem. Theory Comput.}\ }\textbf {\bibinfo {volume} {9}},\ \bibinfo {pages}
  {338} (\bibinfo {year} {2013})}\BibitemShut {NoStop}%
\bibitem [{\citenamefont {Gaus}\ \emph {et~al.}(2014)\citenamefont {Gaus},
  \citenamefont {Lu}, \citenamefont {Elstner},\ and\ \citenamefont
  {Cui}}]{Elstner3obSPParameters2014}%
  \BibitemOpen
  \bibfield  {author} {\bibinfo {author} {\bibfnamefont {M.}~\bibnamefont
  {Gaus}}, \bibinfo {author} {\bibfnamefont {X.}~\bibnamefont {Lu}}, \bibinfo
  {author} {\bibfnamefont {M.}~\bibnamefont {Elstner}}, \ and\ \bibinfo
  {author} {\bibfnamefont {Q.}~\bibnamefont {Cui}},\ }\href {\doibase
  10.1021/ct401002w} {\bibfield  {journal} {\bibinfo  {journal} {J. Chem.
  Theory Comput.}\ }\textbf {\bibinfo {volume} {10}},\ \bibinfo {pages} {1518}
  (\bibinfo {year} {2014})}\BibitemShut {NoStop}%
\bibitem [{\citenamefont {Lu}\ \emph {et~al.}(2015)\citenamefont {Lu},
  \citenamefont {Gaus}, \citenamefont {Elstner},\ and\ \citenamefont
  {Cui}}]{Elstner3obMgZnParameters2015}%
  \BibitemOpen
  \bibfield  {author} {\bibinfo {author} {\bibfnamefont {X.}~\bibnamefont
  {Lu}}, \bibinfo {author} {\bibfnamefont {M.}~\bibnamefont {Gaus}}, \bibinfo
  {author} {\bibfnamefont {M.}~\bibnamefont {Elstner}}, \ and\ \bibinfo
  {author} {\bibfnamefont {Q.}~\bibnamefont {Cui}},\ }\href {\doibase
  10.1021/jp506557r} {\bibfield  {journal} {\bibinfo  {journal} {J. Phys. Chem.
  B}\ }\textbf {\bibinfo {volume} {119}},\ \bibinfo {pages} {1062} (\bibinfo
  {year} {2015})}\BibitemShut {NoStop}%
\bibitem [{\citenamefont {Kubillus}\ \emph {et~al.}(2015)\citenamefont
  {Kubillus}, \citenamefont {Kuba{\v{r}}}, \citenamefont {Gaus}, \citenamefont
  {{\v{R}}ez{\'{a}}{\v{c}}},\ and\ \citenamefont
  {Elstner}}]{Elstner3obMoreParameters2015}%
  \BibitemOpen
  \bibfield  {author} {\bibinfo {author} {\bibfnamefont {M.}~\bibnamefont
  {Kubillus}}, \bibinfo {author} {\bibfnamefont {T.}~\bibnamefont
  {Kuba{\v{r}}}}, \bibinfo {author} {\bibfnamefont {M.}~\bibnamefont {Gaus}},
  \bibinfo {author} {\bibfnamefont {J.}~\bibnamefont
  {{\v{R}}ez{\'{a}}{\v{c}}}}, \ and\ \bibinfo {author} {\bibfnamefont
  {M.}~\bibnamefont {Elstner}},\ }\href {\doibase 10.1021/ct5009137} {\bibfield
   {journal} {\bibinfo  {journal} {J. Chem. Theory Comput.}\ }\textbf {\bibinfo
  {volume} {11}},\ \bibinfo {pages} {332} (\bibinfo {year} {2015})}\BibitemShut
  {NoStop}%
\bibitem [{\citenamefont {Wahiduzzaman}\ \emph {et~al.}(2013)\citenamefont
  {Wahiduzzaman}, \citenamefont {Oliveira}, \citenamefont {Philipsen},
  \citenamefont {Zhechkov}, \citenamefont {van Lenthe}, \citenamefont {Witek},\
  and\ \citenamefont {Heine}}]{WahiduzzamanQuasinano2013}%
  \BibitemOpen
  \bibfield  {author} {\bibinfo {author} {\bibfnamefont {M.}~\bibnamefont
  {Wahiduzzaman}}, \bibinfo {author} {\bibfnamefont {A.~F.}\ \bibnamefont
  {Oliveira}}, \bibinfo {author} {\bibfnamefont {P.}~\bibnamefont {Philipsen}},
  \bibinfo {author} {\bibfnamefont {L.}~\bibnamefont {Zhechkov}}, \bibinfo
  {author} {\bibfnamefont {E.}~\bibnamefont {van Lenthe}}, \bibinfo {author}
  {\bibfnamefont {H.~A.}\ \bibnamefont {Witek}}, \ and\ \bibinfo {author}
  {\bibfnamefont {T.}~\bibnamefont {Heine}},\ }\href {\doibase
  10.1021/ct4004959} {\bibfield  {journal} {\bibinfo  {journal} {J. Chem.
  Theory Comput.}\ }\textbf {\bibinfo {volume} {9}},\ \bibinfo {pages} {4006}
  (\bibinfo {year} {2013})}\BibitemShut {NoStop}%
\bibitem [{\citenamefont {Oliveira}, \citenamefont {Philipsen},\ and\
  \citenamefont {Heine}(2015)}]{OliveiraQUASINANO2015}%
  \BibitemOpen
  \bibfield  {author} {\bibinfo {author} {\bibfnamefont {A.~F.}\ \bibnamefont
  {Oliveira}}, \bibinfo {author} {\bibfnamefont {P.}~\bibnamefont {Philipsen}},
  \ and\ \bibinfo {author} {\bibfnamefont {T.}~\bibnamefont {Heine}},\ }\href
  {\doibase 10.1021/acs.jctc.5b00702} {\bibfield  {journal} {\bibinfo
  {journal} {J. Chem. Theory Comput.}\ }\textbf {\bibinfo {volume} {11}},\
  \bibinfo {pages} {5209} (\bibinfo {year} {2015})}\BibitemShut {NoStop}%
\bibitem [{\citenamefont {Runge}\ and\ \citenamefont
  {Gross}(1984)}]{RungeGrossTheorem1984}%
  \BibitemOpen
  \bibfield  {author} {\bibinfo {author} {\bibfnamefont {E.}~\bibnamefont
  {Runge}}\ and\ \bibinfo {author} {\bibfnamefont {E.~K.~U.}\ \bibnamefont
  {Gross}},\ }\href {\doibase 10.1103/PhysRevLett.52.997} {\bibfield  {journal}
  {\bibinfo  {journal} {Phys. Rev. Lett.}\ }\textbf {\bibinfo {volume} {52}},\
  \bibinfo {pages} {997} (\bibinfo {year} {1984})}\BibitemShut {NoStop}%
\bibitem [{\citenamefont {Casida}(1995)}]{CasidaTDDFT1995}%
  \BibitemOpen
  \bibfield  {author} {\bibinfo {author} {\bibfnamefont {M.~E.}\ \bibnamefont
  {Casida}},\ }in\ \href {\doibase 10.1142/9789812830586_0005} {\emph {\bibinfo
  {booktitle} {Recent Advances in Density Functional Methods}}}\ (\bibinfo
  {year} {1995})\ Chap.~\bibinfo {chapter} {5}, pp.\ \bibinfo {pages}
  {155--192}\BibitemShut {NoStop}%
\bibitem [{\citenamefont {Casida}\ and\ \citenamefont
  {Huix-Rotllant}(2012)}]{CasidaTDDFTReview2012}%
  \BibitemOpen
  \bibfield  {author} {\bibinfo {author} {\bibfnamefont {M.~E.}\ \bibnamefont
  {Casida}}\ and\ \bibinfo {author} {\bibfnamefont {M.}~\bibnamefont
  {Huix-Rotllant}},\ }\href {\doibase 10.1146/annurev-physchem-032511-143803}
  {\bibfield  {journal} {\bibinfo  {journal} {Annu. Rev. Phys. Chem.}\ }\textbf
  {\bibinfo {volume} {63}},\ \bibinfo {pages} {287} (\bibinfo {year}
  {2012})}\BibitemShut {NoStop}%
\bibitem [{\citenamefont {Casida}\ and\ \citenamefont
  {Wesołowski}(2003)}]{CasidaSubsysTDDFT2003}%
  \BibitemOpen
  \bibfield  {author} {\bibinfo {author} {\bibfnamefont {M.~E.}\ \bibnamefont
  {Casida}}\ and\ \bibinfo {author} {\bibfnamefont {T.~A.}\ \bibnamefont
  {Wesołowski}},\ }\href {\doibase 10.1002/qua.10744} {\bibfield  {journal}
  {\bibinfo  {journal} {Int. J. Quantum Chem.}\ }\textbf {\bibinfo {volume}
  {96}},\ \bibinfo {pages} {577} (\bibinfo {year} {2003})}\BibitemShut
  {NoStop}%
\bibitem [{\citenamefont {Neugebauer}(2007)}]{NeugebauerSubsysTDDFT2007}%
  \BibitemOpen
  \bibfield  {author} {\bibinfo {author} {\bibfnamefont {J.}~\bibnamefont
  {Neugebauer}},\ }\href {\doibase 10.1063/1.2713754} {\bibfield  {journal}
  {\bibinfo  {journal} {J. Chem. Phys.}\ }\textbf {\bibinfo {volume} {126}},\
  \bibinfo {pages} {134116} (\bibinfo {year} {2007})}\BibitemShut {NoStop}%
\bibitem [{\citenamefont {Hirata}\ and\ \citenamefont
  {Head-Gordon}(1999)}]{HirataTDA1999}%
  \BibitemOpen
  \bibfield  {author} {\bibinfo {author} {\bibfnamefont {S.}~\bibnamefont
  {Hirata}}\ and\ \bibinfo {author} {\bibfnamefont {M.}~\bibnamefont
  {Head-Gordon}},\ }\href {\doibase
  http://dx.doi.org/10.1016/S0009-2614(99)01149-5} {\bibfield  {journal}
  {\bibinfo  {journal} {Chem. Phys. Lett.}\ }\textbf {\bibinfo {volume}
  {314}},\ \bibinfo {pages} {291} (\bibinfo {year} {1999})}\BibitemShut
  {NoStop}%
\bibitem [{\citenamefont {Grimme}(2013)}]{GrimmeSimplifiedTDA2013}%
  \BibitemOpen
  \bibfield  {author} {\bibinfo {author} {\bibfnamefont {S.}~\bibnamefont
  {Grimme}},\ }\href {\doibase 10.1063/1.4811331} {\bibfield  {journal}
  {\bibinfo  {journal} {J. Chem. Phys.}\ }\textbf {\bibinfo {volume} {138}},\
  \bibinfo {pages} {244104} (\bibinfo {year} {2013})}\BibitemShut {NoStop}%
\bibitem [{\citenamefont {Risthaus}, \citenamefont {Hansen},\ and\
  \citenamefont {Grimme}(2014)}]{GrimmeRangeSepsTDA2014}%
  \BibitemOpen
  \bibfield  {author} {\bibinfo {author} {\bibfnamefont {T.}~\bibnamefont
  {Risthaus}}, \bibinfo {author} {\bibfnamefont {A.}~\bibnamefont {Hansen}}, \
  and\ \bibinfo {author} {\bibfnamefont {S.}~\bibnamefont {Grimme}},\ }\href
  {\doibase 10.1039/C3CP54517B} {\bibfield  {journal} {\bibinfo  {journal}
  {Phys. Chem. Chem. Phys.}\ }\textbf {\bibinfo {volume} {16}},\ \bibinfo
  {pages} {14408} (\bibinfo {year} {2014})}\BibitemShut {NoStop}%
\bibitem [{\citenamefont {R\"uger}\ \emph {et~al.}(2015)\citenamefont
  {R\"uger}, \citenamefont {van Lenthe}, \citenamefont {Lu}, \citenamefont
  {Frenzel}, \citenamefont {Heine},\ and\ \citenamefont
  {Visscher}}]{RugerIntensitySelection2015}%
  \BibitemOpen
  \bibfield  {author} {\bibinfo {author} {\bibfnamefont {R.}~\bibnamefont
  {R\"uger}}, \bibinfo {author} {\bibfnamefont {E.}~\bibnamefont {van Lenthe}},
  \bibinfo {author} {\bibfnamefont {Y.}~\bibnamefont {Lu}}, \bibinfo {author}
  {\bibfnamefont {J.}~\bibnamefont {Frenzel}}, \bibinfo {author} {\bibfnamefont
  {T.}~\bibnamefont {Heine}}, \ and\ \bibinfo {author} {\bibfnamefont
  {L.}~\bibnamefont {Visscher}},\ }\href {\doibase 10.1021/ct500838h}
  {\bibfield  {journal} {\bibinfo  {journal} {J. Chem. Theory Comput.}\
  }\textbf {\bibinfo {volume} {11}},\ \bibinfo {pages} {157} (\bibinfo {year}
  {2015})}\BibitemShut {NoStop}%
\bibitem [{\citenamefont {Bannwarth}\ and\ \citenamefont
  {Grimme}(2014)}]{GrimmeSimplifiedTDDFT2014}%
  \BibitemOpen
  \bibfield  {author} {\bibinfo {author} {\bibfnamefont {C.}~\bibnamefont
  {Bannwarth}}\ and\ \bibinfo {author} {\bibfnamefont {S.}~\bibnamefont
  {Grimme}},\ }\href {\doibase 10.1016/j.comptc.2014.02.023} {\bibfield
  {journal} {\bibinfo  {journal} {Comput. Theor. Chem.}\ }\textbf {\bibinfo
  {volume} {1040--1041}},\ \bibinfo {pages} {45} (\bibinfo {year}
  {2014})}\BibitemShut {NoStop}%
\bibitem [{\citenamefont {Niehaus}\ \emph
  {et~al.}(2001{\natexlab{a}})\citenamefont {Niehaus}, \citenamefont {Suhai},
  \citenamefont {Della~Sala}, \citenamefont {Lugli}, \citenamefont {Elstner},
  \citenamefont {Seifert},\ and\ \citenamefont
  {Frauenheim}}]{NiehausTDDFTB2001}%
  \BibitemOpen
  \bibfield  {author} {\bibinfo {author} {\bibfnamefont {T.~A.}\ \bibnamefont
  {Niehaus}}, \bibinfo {author} {\bibfnamefont {S.}~\bibnamefont {Suhai}},
  \bibinfo {author} {\bibfnamefont {F.}~\bibnamefont {Della~Sala}}, \bibinfo
  {author} {\bibfnamefont {P.}~\bibnamefont {Lugli}}, \bibinfo {author}
  {\bibfnamefont {M.}~\bibnamefont {Elstner}}, \bibinfo {author} {\bibfnamefont
  {G.}~\bibnamefont {Seifert}}, \ and\ \bibinfo {author} {\bibfnamefont
  {T.}~\bibnamefont {Frauenheim}},\ }\href {\doibase
  10.1103/PhysRevB.63.085108} {\bibfield  {journal} {\bibinfo  {journal} {Phys.
  Rev. B}\ }\textbf {\bibinfo {volume} {63}},\ \bibinfo {pages} {085108}
  (\bibinfo {year} {2001}{\natexlab{a}})}\BibitemShut {NoStop}%
\bibitem [{\citenamefont {Domínguez}\ \emph {et~al.}(2013)\citenamefont
  {Domínguez}, \citenamefont {Aradi}, \citenamefont {Frauenheim},
  \citenamefont {Lutsker},\ and\ \citenamefont
  {Niehaus}}]{NiehausTDDFTBOnsiteAndFracOcc2013}%
  \BibitemOpen
  \bibfield  {author} {\bibinfo {author} {\bibfnamefont {A.}~\bibnamefont
  {Domínguez}}, \bibinfo {author} {\bibfnamefont {B.}~\bibnamefont {Aradi}},
  \bibinfo {author} {\bibfnamefont {T.}~\bibnamefont {Frauenheim}}, \bibinfo
  {author} {\bibfnamefont {V.}~\bibnamefont {Lutsker}}, \ and\ \bibinfo
  {author} {\bibfnamefont {T.~A.}\ \bibnamefont {Niehaus}},\ }\href {\doibase
  10.1021/ct400123t} {\bibfield  {journal} {\bibinfo  {journal} {J. Chem.
  Theory Comput.}\ }\textbf {\bibinfo {volume} {9}},\ \bibinfo {pages} {4901}
  (\bibinfo {year} {2013})}\BibitemShut {NoStop}%
\bibitem [{\citenamefont {Trani}\ \emph {et~al.}(2011)\citenamefont {Trani},
  \citenamefont {Scalmani}, \citenamefont {Zheng}, \citenamefont {Carnimeo},
  \citenamefont {Frisch},\ and\ \citenamefont {Barone}}]{GaussianTDDFTB2011}%
  \BibitemOpen
  \bibfield  {author} {\bibinfo {author} {\bibfnamefont {F.}~\bibnamefont
  {Trani}}, \bibinfo {author} {\bibfnamefont {G.}~\bibnamefont {Scalmani}},
  \bibinfo {author} {\bibfnamefont {G.}~\bibnamefont {Zheng}}, \bibinfo
  {author} {\bibfnamefont {I.}~\bibnamefont {Carnimeo}}, \bibinfo {author}
  {\bibfnamefont {M.~J.}\ \bibnamefont {Frisch}}, \ and\ \bibinfo {author}
  {\bibfnamefont {V.}~\bibnamefont {Barone}},\ }\href {\doibase
  10.1021/ct200461y} {\bibfield  {journal} {\bibinfo  {journal} {J. Chem.
  Theory Comput.}\ }\textbf {\bibinfo {volume} {7}},\ \bibinfo {pages} {3304}
  (\bibinfo {year} {2011})}\BibitemShut {NoStop}%
\bibitem [{\citenamefont {Joswig}\ \emph {et~al.}(2003)\citenamefont {Joswig},
  \citenamefont {Seifert}, \citenamefont {Niehaus},\ and\ \citenamefont
  {Springborg}}]{tddftbapp1_doi:10.1021/jp026752s}%
  \BibitemOpen
  \bibfield  {author} {\bibinfo {author} {\bibfnamefont {J.-O.}\ \bibnamefont
  {Joswig}}, \bibinfo {author} {\bibfnamefont {G.}~\bibnamefont {Seifert}},
  \bibinfo {author} {\bibfnamefont {T.~A.}\ \bibnamefont {Niehaus}}, \ and\
  \bibinfo {author} {\bibfnamefont {M.}~\bibnamefont {Springborg}},\ }\href
  {\doibase 10.1021/jp026752s} {\bibfield  {journal} {\bibinfo  {journal} {J.
  Phys. Chem. B}\ }\textbf {\bibinfo {volume} {107}},\ \bibinfo {pages} {2897}
  (\bibinfo {year} {2003})}\BibitemShut {NoStop}%
\bibitem [{\citenamefont {Goswami}\ \emph {et~al.}(2006)\citenamefont
  {Goswami}, \citenamefont {Pal}, \citenamefont {Sarkar}, \citenamefont
  {Seifert},\ and\ \citenamefont {Springborg}}]{tddftbapp2_PhysRevB.73.205312}%
  \BibitemOpen
  \bibfield  {author} {\bibinfo {author} {\bibfnamefont {B.}~\bibnamefont
  {Goswami}}, \bibinfo {author} {\bibfnamefont {S.}~\bibnamefont {Pal}},
  \bibinfo {author} {\bibfnamefont {P.}~\bibnamefont {Sarkar}}, \bibinfo
  {author} {\bibfnamefont {G.}~\bibnamefont {Seifert}}, \ and\ \bibinfo
  {author} {\bibfnamefont {M.}~\bibnamefont {Springborg}},\ }\href {\doibase
  10.1103/PhysRevB.73.205312} {\bibfield  {journal} {\bibinfo  {journal} {Phys.
  Rev. B}\ }\textbf {\bibinfo {volume} {73}},\ \bibinfo {pages} {205312}
  (\bibinfo {year} {2006})}\BibitemShut {NoStop}%
\bibitem [{\citenamefont {Frenzel}, \citenamefont {Joswig},\ and\ \citenamefont
  {Seifert}(2007)}]{tddftbapp3_doi:10.1021/jp071125u}%
  \BibitemOpen
  \bibfield  {author} {\bibinfo {author} {\bibfnamefont {J.}~\bibnamefont
  {Frenzel}}, \bibinfo {author} {\bibfnamefont {J.-O.}\ \bibnamefont {Joswig}},
  \ and\ \bibinfo {author} {\bibfnamefont {G.}~\bibnamefont {Seifert}},\ }\href
  {\doibase 10.1021/jp071125u} {\bibfield  {journal} {\bibinfo  {journal} {J.
  Phys. Chem. C}\ }\textbf {\bibinfo {volume} {111}},\ \bibinfo {pages} {10761}
  (\bibinfo {year} {2007})}\BibitemShut {NoStop}%
\bibitem [{\citenamefont {Li}\ \emph {et~al.}(2007)\citenamefont {Li},
  \citenamefont {Zhang}, \citenamefont {Niehaus}, \citenamefont {Frauenheim},\
  and\ \citenamefont {Lee}}]{tddftbapp4_doi:10.1021/ct700041v}%
  \BibitemOpen
  \bibfield  {author} {\bibinfo {author} {\bibfnamefont {Q.~S.}\ \bibnamefont
  {Li}}, \bibinfo {author} {\bibfnamefont {R.~Q.}\ \bibnamefont {Zhang}},
  \bibinfo {author} {\bibfnamefont {T.~A.}\ \bibnamefont {Niehaus}}, \bibinfo
  {author} {\bibfnamefont {T.}~\bibnamefont {Frauenheim}}, \ and\ \bibinfo
  {author} {\bibfnamefont {S.~T.}\ \bibnamefont {Lee}},\ }\href {\doibase
  10.1021/ct700041v} {\bibfield  {journal} {\bibinfo  {journal} {J. Chem.
  Theory Comput.}\ }\textbf {\bibinfo {volume} {3}},\ \bibinfo {pages} {1518}
  (\bibinfo {year} {2007})}\BibitemShut {NoStop}%
\bibitem [{\citenamefont {Wang}\ \emph
  {et~al.}(2007{\natexlab{a}})\citenamefont {Wang}, \citenamefont {Zhang},
  \citenamefont {Niehaus},\ and\ \citenamefont
  {Frauenheim}}]{tddftbapp5_doi:10.1021/jp065704v}%
  \BibitemOpen
  \bibfield  {author} {\bibinfo {author} {\bibfnamefont {X.}~\bibnamefont
  {Wang}}, \bibinfo {author} {\bibnamefont {Zhang}}, \bibinfo {author}
  {\bibnamefont {Niehaus}}, \ and\ \bibinfo {author} {\bibfnamefont
  {T.}~\bibnamefont {Frauenheim}},\ }\href
  {http://pubs.acs.org/doi/abs/10.1021/jp065704v} {\bibfield  {journal}
  {\bibinfo  {journal} {J. Phys. Chem. C}\ }\textbf {\bibinfo {volume} {111}},\
  \bibinfo {pages} {2394} (\bibinfo {year} {2007}{\natexlab{a}})}\BibitemShut
  {NoStop}%
\bibitem [{\citenamefont {Wang}\ \emph
  {et~al.}(2007{\natexlab{b}})\citenamefont {Wang}, \citenamefont {Zhang},
  \citenamefont {Lee}, \citenamefont {Niehaus},\ and\ \citenamefont
  {Frauenheim}}]{tddftbapp6_10.1063/1.2715101}%
  \BibitemOpen
  \bibfield  {author} {\bibinfo {author} {\bibfnamefont {X.}~\bibnamefont
  {Wang}}, \bibinfo {author} {\bibfnamefont {R.~Q.}\ \bibnamefont {Zhang}},
  \bibinfo {author} {\bibfnamefont {S.~T.}\ \bibnamefont {Lee}}, \bibinfo
  {author} {\bibfnamefont {T.~A.}\ \bibnamefont {Niehaus}}, \ and\ \bibinfo
  {author} {\bibfnamefont {T.}~\bibnamefont {Frauenheim}},\ }\href
  {http://scitation.aip.org/content/aip/journal/apl/90/12/10.1063/1.2715101}
  {\bibfield  {journal} {\bibinfo  {journal} {Appl. Phys. Lett.}\ }\textbf
  {\bibinfo {volume} {90}},\ \bibinfo {eid} {123116} (\bibinfo {year}
  {2007}{\natexlab{b}})}\BibitemShut {NoStop}%
\bibitem [{\citenamefont {Li}\ \emph {et~al.}(2008)\citenamefont {Li},
  \citenamefont {Zhang}, \citenamefont {Lee}, \citenamefont {Niehaus},\ and\
  \citenamefont {Frauenheim}}]{tddftbapp7_doi:10.1063/1.2940735}%
  \BibitemOpen
  \bibfield  {author} {\bibinfo {author} {\bibfnamefont {Q.~S.}\ \bibnamefont
  {Li}}, \bibinfo {author} {\bibfnamefont {R.~Q.}\ \bibnamefont {Zhang}},
  \bibinfo {author} {\bibfnamefont {S.~T.}\ \bibnamefont {Lee}}, \bibinfo
  {author} {\bibfnamefont {T.~A.}\ \bibnamefont {Niehaus}}, \ and\ \bibinfo
  {author} {\bibfnamefont {T.}~\bibnamefont {Frauenheim}},\ }\href
  {http://scitation.aip.org/content/aip/journal/jcp/128/24/10.1063/1.2940735}
  {\bibfield  {journal} {\bibinfo  {journal} {J. Chem. Phys.}\ }\textbf
  {\bibinfo {volume} {128}},\ \bibinfo {eid} {244714} (\bibinfo {year}
  {2008})}\BibitemShut {NoStop}%
\bibitem [{\citenamefont {Mitrić}\ \emph {et~al.}(2009)\citenamefont
  {Mitrić}, \citenamefont {Werner}, \citenamefont {Wohlgemuth}, \citenamefont
  {Seifert},\ and\ \citenamefont
  {Bonačić-Koutecký}}]{BonacicNonAdMDWithTDDFTB2009}%
  \BibitemOpen
  \bibfield  {author} {\bibinfo {author} {\bibfnamefont {R.}~\bibnamefont
  {Mitrić}}, \bibinfo {author} {\bibfnamefont {U.}~\bibnamefont {Werner}},
  \bibinfo {author} {\bibfnamefont {M.}~\bibnamefont {Wohlgemuth}}, \bibinfo
  {author} {\bibfnamefont {G.}~\bibnamefont {Seifert}}, \ and\ \bibinfo
  {author} {\bibfnamefont {V.}~\bibnamefont {Bonačić-Koutecký}},\ }\href
  {\doibase 10.1021/jp905600w} {\bibfield  {journal} {\bibinfo  {journal} {J.
  Phys. Chem. A}\ }\textbf {\bibinfo {volume} {113}},\ \bibinfo {pages} {12700}
  (\bibinfo {year} {2009})}\BibitemShut {NoStop}%
\bibitem [{\citenamefont {Zhang}\ \emph {et~al.}(2012)\citenamefont {Zhang},
  \citenamefont {De~Sarkar}, \citenamefont {Niehaus},\ and\ \citenamefont
  {Frauenheim}}]{tddftbapp9_PSSB:PSSB201100719}%
  \BibitemOpen
  \bibfield  {author} {\bibinfo {author} {\bibfnamefont {R.-Q.}\ \bibnamefont
  {Zhang}}, \bibinfo {author} {\bibfnamefont {A.}~\bibnamefont {De~Sarkar}},
  \bibinfo {author} {\bibfnamefont {T.~A.}\ \bibnamefont {Niehaus}}, \ and\
  \bibinfo {author} {\bibfnamefont {T.}~\bibnamefont {Frauenheim}},\ }\href
  {http://dx.doi.org/10.1002/pssb.201100719} {\bibfield  {journal} {\bibinfo
  {journal} {Phys. Status Solidi B}\ }\textbf {\bibinfo {volume} {249}},\
  \bibinfo {pages} {401} (\bibinfo {year} {2012})}\BibitemShut {NoStop}%
\bibitem [{\citenamefont {Fan}\ \emph {et~al.}(2014)\citenamefont {Fan},
  \citenamefont {Li}, \citenamefont {Liu},\ and\ \citenamefont
  {He}}]{tddftbapp10_Fan201417}%
  \BibitemOpen
  \bibfield  {author} {\bibinfo {author} {\bibfnamefont {G.-H.}\ \bibnamefont
  {Fan}}, \bibinfo {author} {\bibfnamefont {X.}~\bibnamefont {Li}}, \bibinfo
  {author} {\bibfnamefont {J.-Y.}\ \bibnamefont {Liu}}, \ and\ \bibinfo
  {author} {\bibfnamefont {G.-Z.}\ \bibnamefont {He}},\ }\href {\doibase
  http://dx.doi.org/10.1016/j.comptc.2013.12.010} {\bibfield  {journal}
  {\bibinfo  {journal} {Comp. Theor. Chem.}\ }\textbf {\bibinfo {volume}
  {1030}},\ \bibinfo {pages} {17} (\bibinfo {year} {2014})}\BibitemShut
  {NoStop}%
\bibitem [{\citenamefont {Niehaus}(2009)}]{NiehausTDDFTBReview2009}%
  \BibitemOpen
  \bibfield  {author} {\bibinfo {author} {\bibfnamefont {T.~A.}\ \bibnamefont
  {Niehaus}},\ }\href {\doibase 10.1016/j.theochem.2009.04.034} {\bibfield
  {journal} {\bibinfo  {journal} {J. Mol. Struc.: THEOCHEM}\ }\textbf {\bibinfo
  {volume} {914}},\ \bibinfo {pages} {38} (\bibinfo {year} {2009})}\BibitemShut
  {NoStop}%
\bibitem [{\citenamefont {Becke}(1993)}]{BeckeHybrids1993}%
  \BibitemOpen
  \bibfield  {author} {\bibinfo {author} {\bibfnamefont {A.~D.}\ \bibnamefont
  {Becke}},\ }\href {\doibase 10.1063/1.464304} {\bibfield  {journal} {\bibinfo
   {journal} {J. Chem. Phys.}\ }\textbf {\bibinfo {volume} {98}},\ \bibinfo
  {pages} {1372} (\bibinfo {year} {1993})}\BibitemShut {NoStop}%
\bibitem [{\citenamefont {Baerends}, \citenamefont {Gritsenko},\ and\
  \citenamefont {van Meer}(2013)}]{BaerendsKSOrbitals2013}%
  \BibitemOpen
  \bibfield  {author} {\bibinfo {author} {\bibfnamefont {E.~J.}\ \bibnamefont
  {Baerends}}, \bibinfo {author} {\bibfnamefont {O.~V.}\ \bibnamefont
  {Gritsenko}}, \ and\ \bibinfo {author} {\bibfnamefont {R.}~\bibnamefont {van
  Meer}},\ }\href {\doibase 10.1039/c3cp52547c} {\bibfield  {journal} {\bibinfo
   {journal} {Phys. Chem. Chem. Phys.}\ }\textbf {\bibinfo {volume} {15}},\
  \bibinfo {pages} {16408} (\bibinfo {year} {2013})}\BibitemShut {NoStop}%
\bibitem [{\citenamefont {van Meer}, \citenamefont {Gritsenko},\ and\
  \citenamefont {Baerends}(2014)}]{BaerendsKSOrbitals2014}%
  \BibitemOpen
  \bibfield  {author} {\bibinfo {author} {\bibfnamefont {R.}~\bibnamefont {van
  Meer}}, \bibinfo {author} {\bibfnamefont {O.~V.}\ \bibnamefont {Gritsenko}},
  \ and\ \bibinfo {author} {\bibfnamefont {E.~J.}\ \bibnamefont {Baerends}},\
  }\href {\doibase 10.1021/ct500727c} {\bibfield  {journal} {\bibinfo
  {journal} {J. Chem. Theory Comput.}\ }\textbf {\bibinfo {volume} {10}},\
  \bibinfo {pages} {4432} (\bibinfo {year} {2014})}\BibitemShut {NoStop}%
\bibitem [{\citenamefont {Mulliken}(1939)}]{MullikenIntensities1939}%
  \BibitemOpen
  \bibfield  {author} {\bibinfo {author} {\bibfnamefont {R.~S.}\ \bibnamefont
  {Mulliken}},\ }\href {\doibase 10.1063/1.1750317} {\bibfield  {journal}
  {\bibinfo  {journal} {J. Chem. Phys.}\ }\textbf {\bibinfo {volume} {7}},\
  \bibinfo {pages} {14} (\bibinfo {year} {1939})}\BibitemShut {NoStop}%
\bibitem [{\citenamefont {Marques}\ \emph {et~al.}(2012)\citenamefont
  {Marques}, \citenamefont {Maitra}, \citenamefont {Nogueira}, \citenamefont
  {Gross},\ and\ \citenamefont {Rubio}}]{TDDFTBook2012}%
  \BibitemOpen
  \bibinfo {editor} {\bibfnamefont {M.~A.}\ \bibnamefont {Marques}}, \bibinfo
  {editor} {\bibfnamefont {N.~T.}\ \bibnamefont {Maitra}}, \bibinfo {editor}
  {\bibfnamefont {F.~M.}\ \bibnamefont {Nogueira}}, \bibinfo {editor}
  {\bibfnamefont {E.}~\bibnamefont {Gross}}, \ and\ \bibinfo {editor}
  {\bibfnamefont {A.}~\bibnamefont {Rubio}},\ eds.,\ \href {\doibase
  10.1007/978-3-642-23518-4} {\emph {\bibinfo {title} {Fundamentals of
  Time-Dependent Density Functional Theory}}}\ (\bibinfo  {publisher} {Springer
  Berlin Heidelberg},\ \bibinfo {year} {2012})\BibitemShut {NoStop}%
\bibitem [{\citenamefont {van Gisbergen}, \citenamefont {Snijders},\ and\
  \citenamefont {Baerends}(1999)}]{GisbergenADFTDDFTB1999}%
  \BibitemOpen
  \bibfield  {author} {\bibinfo {author} {\bibfnamefont {S.~J.~A.}\
  \bibnamefont {van Gisbergen}}, \bibinfo {author} {\bibfnamefont
  {J.}~\bibnamefont {Snijders}}, \ and\ \bibinfo {author} {\bibfnamefont
  {E.~J.}\ \bibnamefont {Baerends}},\ }\href {\doibase
  http://dx.doi.org/10.1016/S0010-4655(99)00187-3} {\bibfield  {journal}
  {\bibinfo  {journal} {Comput. Phys. Commun.}\ }\textbf {\bibinfo {volume}
  {118}},\ \bibinfo {pages} {119} (\bibinfo {year} {1999})}\BibitemShut
  {NoStop}%
\bibitem [{\citenamefont {Mulliken}(1955)}]{MullikenPopulationAnalysis1955}%
  \BibitemOpen
  \bibfield  {author} {\bibinfo {author} {\bibfnamefont {R.~S.}\ \bibnamefont
  {Mulliken}},\ }\href {\doibase 10.1063/1.1740588} {\bibfield  {journal}
  {\bibinfo  {journal} {J. Chem. Phys.}\ }\textbf {\bibinfo {volume} {23}},\
  \bibinfo {pages} {1833} (\bibinfo {year} {1955})}\BibitemShut {NoStop}%
\bibitem [{\citenamefont {Mineva}\ and\ \citenamefont
  {Heine}(2004)}]{MinevaHeine1}%
  \BibitemOpen
  \bibfield  {author} {\bibinfo {author} {\bibfnamefont {T.}~\bibnamefont
  {Mineva}}\ and\ \bibinfo {author} {\bibfnamefont {T.}~\bibnamefont {Heine}},\
  }\href {\doibase 10.1021/jp048000z} {\bibfield  {journal} {\bibinfo
  {journal} {J. Phys. Chem. A}\ }\textbf {\bibinfo {volume} {108}},\ \bibinfo
  {pages} {11086} (\bibinfo {year} {2004})}\BibitemShut {NoStop}%
\bibitem [{\citenamefont {Mineva}\ and\ \citenamefont
  {Heine}(2006)}]{MinevaHeine2}%
  \BibitemOpen
  \bibfield  {author} {\bibinfo {author} {\bibfnamefont {T.}~\bibnamefont
  {Mineva}}\ and\ \bibinfo {author} {\bibfnamefont {T.}~\bibnamefont {Heine}},\
  }\href {\doibase 10.1002/qua.20897} {\bibfield  {journal} {\bibinfo
  {journal} {Int. J. Quantum Chem.}\ }\textbf {\bibinfo {volume} {106}},\
  \bibinfo {pages} {1396} (\bibinfo {year} {2006})}\BibitemShut {NoStop}%
\bibitem [{\citenamefont {Ghosh}\ and\ \citenamefont
  {Islam}(2009)}]{GhoshIslamChemicalHardness2009}%
  \BibitemOpen
  \bibfield  {author} {\bibinfo {author} {\bibfnamefont {D.~C.}\ \bibnamefont
  {Ghosh}}\ and\ \bibinfo {author} {\bibfnamefont {N.}~\bibnamefont {Islam}},\
  }\href {\doibase 10.1002/qua.22202} {\bibfield  {journal} {\bibinfo
  {journal} {Int. J. Quantum Chem.}\ }\textbf {\bibinfo {volume} {110}},\
  \bibinfo {pages} {1206} (\bibinfo {year} {2009})}\BibitemShut {NoStop}%
\bibitem [{\citenamefont {L\"owdin}(1950)}]{LowdinPopulationAnalysis1950}%
  \BibitemOpen
  \bibfield  {author} {\bibinfo {author} {\bibfnamefont {P.-O.}\ \bibnamefont
  {L\"owdin}},\ }\href {\doibase 10.1063/1.1747632} {\bibfield  {journal}
  {\bibinfo  {journal} {J. Chem. Phys.}\ }\textbf {\bibinfo {volume} {18}},\
  \bibinfo {pages} {365} (\bibinfo {year} {1950})}\BibitemShut {NoStop}%
\bibitem [{\citenamefont {Heine}\ \emph {et~al.}(1999)\citenamefont {Heine},
  \citenamefont {Seifert}, \citenamefont {Fowler},\ and\ \citenamefont
  {Zerbetto}}]{HeineDFTBFullereneNMR1999}%
  \BibitemOpen
  \bibfield  {author} {\bibinfo {author} {\bibfnamefont {T.}~\bibnamefont
  {Heine}}, \bibinfo {author} {\bibfnamefont {G.}~\bibnamefont {Seifert}},
  \bibinfo {author} {\bibfnamefont {P.~W.}\ \bibnamefont {Fowler}}, \ and\
  \bibinfo {author} {\bibfnamefont {F.}~\bibnamefont {Zerbetto}},\ }\href
  {\doibase 10.1021/jp9923062} {\bibfield  {journal} {\bibinfo  {journal} {J.
  Phys. Chem. A}\ }\textbf {\bibinfo {volume} {103}},\ \bibinfo {pages} {8738}
  (\bibinfo {year} {1999})}\BibitemShut {NoStop}%
\bibitem [{\citenamefont {Gritsenko}\ and\ \citenamefont
  {Baerends}(2004)}]{GritsenkoCTFailure2004}%
  \BibitemOpen
  \bibfield  {author} {\bibinfo {author} {\bibfnamefont {O.}~\bibnamefont
  {Gritsenko}}\ and\ \bibinfo {author} {\bibfnamefont {E.~J.}\ \bibnamefont
  {Baerends}},\ }\href {\doibase http://dx.doi.org/10.1063/1.1759320}
  {\bibfield  {journal} {\bibinfo  {journal} {J. Chem. Phys.}\ }\textbf
  {\bibinfo {volume} {121}},\ \bibinfo {pages} {655} (\bibinfo {year}
  {2004})}\BibitemShut {NoStop}%
\bibitem [{\citenamefont {Bredas}(2014)}]{BredasMindTheGap2014}%
  \BibitemOpen
  \bibfield  {author} {\bibinfo {author} {\bibfnamefont {J.-L.}\ \bibnamefont
  {Bredas}},\ }\href {\doibase 10.1039/c3mh00098b} {\bibfield  {journal}
  {\bibinfo  {journal} {Mater. Horiz.}\ }\textbf {\bibinfo {volume} {1}},\
  \bibinfo {pages} {17} (\bibinfo {year} {2014})}\BibitemShut {NoStop}%
\bibitem [{\citenamefont {Iikura}\ \emph {et~al.}(2001)\citenamefont {Iikura},
  \citenamefont {Tsuneda}, \citenamefont {Yanai},\ and\ \citenamefont
  {Hirao}}]{IikuraRangeSep2001}%
  \BibitemOpen
  \bibfield  {author} {\bibinfo {author} {\bibfnamefont {H.}~\bibnamefont
  {Iikura}}, \bibinfo {author} {\bibfnamefont {T.}~\bibnamefont {Tsuneda}},
  \bibinfo {author} {\bibfnamefont {T.}~\bibnamefont {Yanai}}, \ and\ \bibinfo
  {author} {\bibfnamefont {K.}~\bibnamefont {Hirao}},\ }\href {\doibase
  10.1063/1.1383587} {\bibfield  {journal} {\bibinfo  {journal} {J. Chem.
  Phys.}\ }\textbf {\bibinfo {volume} {115}},\ \bibinfo {pages} {3540}
  (\bibinfo {year} {2001})}\BibitemShut {NoStop}%
\bibitem [{\citenamefont {Henderson}, \citenamefont {Janesko},\ and\
  \citenamefont {Scuseria}(2008)}]{HendersonRangeSep2008}%
  \BibitemOpen
  \bibfield  {author} {\bibinfo {author} {\bibfnamefont {T.~M.}\ \bibnamefont
  {Henderson}}, \bibinfo {author} {\bibfnamefont {B.~G.}\ \bibnamefont
  {Janesko}}, \ and\ \bibinfo {author} {\bibfnamefont {G.~E.}\ \bibnamefont
  {Scuseria}},\ }\href {\doibase 10.1021/jp806573k} {\bibfield  {journal}
  {\bibinfo  {journal} {J. Phys. Chem. A}\ }\textbf {\bibinfo {volume} {112}},\
  \bibinfo {pages} {12530} (\bibinfo {year} {2008})}\BibitemShut {NoStop}%
\bibitem [{\citenamefont {Baer}, \citenamefont {Livshits},\ and\ \citenamefont
  {Salzner}(2010)}]{BaerRangeSepReview2010}%
  \BibitemOpen
  \bibfield  {author} {\bibinfo {author} {\bibfnamefont {R.}~\bibnamefont
  {Baer}}, \bibinfo {author} {\bibfnamefont {E.}~\bibnamefont {Livshits}}, \
  and\ \bibinfo {author} {\bibfnamefont {U.}~\bibnamefont {Salzner}},\ }\href
  {\doibase 10.1146/annurev.physchem.012809.103321} {\bibfield  {journal}
  {\bibinfo  {journal} {Annual Review of Physical Chemistry}\ }\textbf
  {\bibinfo {volume} {61}},\ \bibinfo {pages} {85} (\bibinfo {year}
  {2010})}\BibitemShut {NoStop}%
\bibitem [{\citenamefont {Zhang}\ and\ \citenamefont
  {Musgrave}(2007)}]{ZhangDFTOrbitalEigenvalues2007}%
  \BibitemOpen
  \bibfield  {author} {\bibinfo {author} {\bibfnamefont {G.}~\bibnamefont
  {Zhang}}\ and\ \bibinfo {author} {\bibfnamefont {C.~B.}\ \bibnamefont
  {Musgrave}},\ }\href {\doibase 10.1021/jp061633o} {\bibfield  {journal}
  {\bibinfo  {journal} {J. Phys. Chem. A}\ }\textbf {\bibinfo {volume} {111}},\
  \bibinfo {pages} {1554} (\bibinfo {year} {2007})}\BibitemShut {NoStop}%
\bibitem [{\citenamefont {Schreiber}\ \emph {et~al.}(2008)\citenamefont
  {Schreiber}, \citenamefont {Silva-Junior}, \citenamefont {Sauer},\ and\
  \citenamefont {Thiel}}]{ThielTestset2008}%
  \BibitemOpen
  \bibfield  {author} {\bibinfo {author} {\bibfnamefont {M.}~\bibnamefont
  {Schreiber}}, \bibinfo {author} {\bibfnamefont {M.~R.}\ \bibnamefont
  {Silva-Junior}}, \bibinfo {author} {\bibfnamefont {S.~P.~A.}\ \bibnamefont
  {Sauer}}, \ and\ \bibinfo {author} {\bibfnamefont {W.}~\bibnamefont
  {Thiel}},\ }\href {\doibase 10.1063/1.2889385} {\bibfield  {journal}
  {\bibinfo  {journal} {J. Chem. Phys.}\ }\textbf {\bibinfo {volume} {128}},\
  \bibinfo {pages} {134110} (\bibinfo {year} {2008})}\BibitemShut {NoStop}%
\bibitem [{\citenamefont {Niehaus}\ \emph
  {et~al.}(2001{\natexlab{b}})\citenamefont {Niehaus}, \citenamefont {Elstner},
  \citenamefont {Frauenheim},\ and\ \citenamefont
  {Suhai}}]{NiehausMIOSulfur2001}%
  \BibitemOpen
  \bibfield  {author} {\bibinfo {author} {\bibfnamefont {T.~A.}\ \bibnamefont
  {Niehaus}}, \bibinfo {author} {\bibfnamefont {M.}~\bibnamefont {Elstner}},
  \bibinfo {author} {\bibfnamefont {T.}~\bibnamefont {Frauenheim}}, \ and\
  \bibinfo {author} {\bibfnamefont {S.}~\bibnamefont {Suhai}},\ }\href
  {\doibase 10.1016/s0166-1280(00)00762-4} {\bibfield  {journal} {\bibinfo
  {journal} {J. Mol. Struc. ({THEOCHEM})}\ }\textbf {\bibinfo {volume} {541}},\
  \bibinfo {pages} {185} (\bibinfo {year} {2001}{\natexlab{b}})}\BibitemShut
  {NoStop}%
\bibitem [{\citenamefont {Yang}\ \emph {et~al.}(2008)\citenamefont {Yang},
  \citenamefont {Yu}, \citenamefont {York}, \citenamefont {Elstner},\ and\
  \citenamefont {Cui}}]{YangMIOPhosphorus2008}%
  \BibitemOpen
  \bibfield  {author} {\bibinfo {author} {\bibfnamefont {Y.}~\bibnamefont
  {Yang}}, \bibinfo {author} {\bibfnamefont {H.}~\bibnamefont {Yu}}, \bibinfo
  {author} {\bibfnamefont {D.}~\bibnamefont {York}}, \bibinfo {author}
  {\bibfnamefont {M.}~\bibnamefont {Elstner}}, \ and\ \bibinfo {author}
  {\bibfnamefont {Q.}~\bibnamefont {Cui}},\ }\href {\doibase 10.1021/ct800330d}
  {\bibfield  {journal} {\bibinfo  {journal} {J. Chem. Theory Comput.}\
  }\textbf {\bibinfo {volume} {4}},\ \bibinfo {pages} {2067} (\bibinfo {year}
  {2008})}\BibitemShut {NoStop}%
\bibitem [{\citenamefont {Perdew}, \citenamefont {Burke},\ and\ \citenamefont
  {Ernzerhof}(1996)}]{PerdewBurkeErnzerhofPBEXCFunc1996}%
  \BibitemOpen
  \bibfield  {author} {\bibinfo {author} {\bibfnamefont {J.~P.}\ \bibnamefont
  {Perdew}}, \bibinfo {author} {\bibfnamefont {K.}~\bibnamefont {Burke}}, \
  and\ \bibinfo {author} {\bibfnamefont {M.}~\bibnamefont {Ernzerhof}},\ }\href
  {\doibase 10.1103/PhysRevLett.77.3865} {\bibfield  {journal} {\bibinfo
  {journal} {Phys. Rev. Lett.}\ }\textbf {\bibinfo {volume} {77}},\ \bibinfo
  {pages} {3865} (\bibinfo {year} {1996})}\BibitemShut {NoStop}%
\bibitem [{\citenamefont {Huix-Rotllant}\ \emph {et~al.}(2011)\citenamefont
  {Huix-Rotllant}, \citenamefont {Ipatov}, \citenamefont {Rubio},\ and\
  \citenamefont {Casida}}]{Huix_Rotllant_2011}%
  \BibitemOpen
  \bibfield  {author} {\bibinfo {author} {\bibfnamefont {M.}~\bibnamefont
  {Huix-Rotllant}}, \bibinfo {author} {\bibfnamefont {A.}~\bibnamefont
  {Ipatov}}, \bibinfo {author} {\bibfnamefont {A.}~\bibnamefont {Rubio}}, \
  and\ \bibinfo {author} {\bibfnamefont {M.~E.}\ \bibnamefont {Casida}},\
  }\href {\doibase 10.1016/j.chemphys.2011.03.019} {\bibfield  {journal}
  {\bibinfo  {journal} {Chem. Phys.}\ }\textbf {\bibinfo {volume} {391}},\
  \bibinfo {pages} {120} (\bibinfo {year} {2011})}\BibitemShut {NoStop}%
\bibitem [{\citenamefont {Casida}\ and\ \citenamefont
  {Huix-Rotllant}(2015)}]{Casida_2015}%
  \BibitemOpen
  \bibfield  {author} {\bibinfo {author} {\bibfnamefont {M.~E.}\ \bibnamefont
  {Casida}}\ and\ \bibinfo {author} {\bibfnamefont {M.}~\bibnamefont
  {Huix-Rotllant}},\ }in\ \href {\doibase 10.1007/128_2015_632} {\emph
  {\bibinfo {booktitle} {Density-Functional Methods for Excited States}}}\
  (\bibinfo  {publisher} {Springer Science + Business Media},\ \bibinfo {year}
  {2015})\ pp.\ \bibinfo {pages} {1--60}\BibitemShut {NoStop}%
\bibitem [{\citenamefont {Casida}\ \emph {et~al.}(1998)\citenamefont {Casida},
  \citenamefont {Jamorski}, \citenamefont {Casida},\ and\ \citenamefont
  {Salahub}}]{CasidaTDDFTLDALimits1998}%
  \BibitemOpen
  \bibfield  {author} {\bibinfo {author} {\bibfnamefont {M.~E.}\ \bibnamefont
  {Casida}}, \bibinfo {author} {\bibfnamefont {C.}~\bibnamefont {Jamorski}},
  \bibinfo {author} {\bibfnamefont {K.~C.}\ \bibnamefont {Casida}}, \ and\
  \bibinfo {author} {\bibfnamefont {D.~R.}\ \bibnamefont {Salahub}},\ }\href
  {\doibase 10.1063/1.475855} {\bibfield  {journal} {\bibinfo  {journal} {J.
  Chem. Phys.}\ }\textbf {\bibinfo {volume} {108}},\ \bibinfo {pages} {4439}
  (\bibinfo {year} {1998})}\BibitemShut {NoStop}%
\bibitem [{\citenamefont {Yasumatsu}\ \emph {et~al.}(1996)\citenamefont
  {Yasumatsu}, \citenamefont {Kondow}, \citenamefont {Kitagawa}, \citenamefont
  {Tabayashi},\ and\ \citenamefont {Shobatake}}]{YasumatsuC60Abs1996}%
  \BibitemOpen
  \bibfield  {author} {\bibinfo {author} {\bibfnamefont {H.}~\bibnamefont
  {Yasumatsu}}, \bibinfo {author} {\bibfnamefont {T.}~\bibnamefont {Kondow}},
  \bibinfo {author} {\bibfnamefont {H.}~\bibnamefont {Kitagawa}}, \bibinfo
  {author} {\bibfnamefont {K.}~\bibnamefont {Tabayashi}}, \ and\ \bibinfo
  {author} {\bibfnamefont {K.}~\bibnamefont {Shobatake}},\ }\href {\doibase
  10.1063/1.470813} {\bibfield  {journal} {\bibinfo  {journal} {J. Chem.
  Phys.}\ }\textbf {\bibinfo {volume} {104}},\ \bibinfo {pages} {899} (\bibinfo
  {year} {1996})}\BibitemShut {NoStop}%
\bibitem [{\citenamefont {Abrefah}\ \emph {et~al.}(1992)\citenamefont
  {Abrefah}, \citenamefont {Olander}, \citenamefont {Balooch},\ and\
  \citenamefont {Siekhaus}}]{AbrefahC60VaporPressure1992}%
  \BibitemOpen
  \bibfield  {author} {\bibinfo {author} {\bibfnamefont {J.}~\bibnamefont
  {Abrefah}}, \bibinfo {author} {\bibfnamefont {D.~R.}\ \bibnamefont
  {Olander}}, \bibinfo {author} {\bibfnamefont {M.}~\bibnamefont {Balooch}}, \
  and\ \bibinfo {author} {\bibfnamefont {W.~J.}\ \bibnamefont {Siekhaus}},\
  }\href {\doibase 10.1063/1.107327} {\bibfield  {journal} {\bibinfo  {journal}
  {Appl. Phys. Lett.}\ }\textbf {\bibinfo {volume} {60}},\ \bibinfo {pages}
  {1313} (\bibinfo {year} {1992})}\BibitemShut {NoStop}%
\bibitem [{\citenamefont {Strain}, \citenamefont {Thomas},\ and\ \citenamefont
  {Katz}(1963)}]{StrainChlA1963}%
  \BibitemOpen
  \bibfield  {author} {\bibinfo {author} {\bibfnamefont {H.~H.}\ \bibnamefont
  {Strain}}, \bibinfo {author} {\bibfnamefont {M.~R.}\ \bibnamefont {Thomas}},
  \ and\ \bibinfo {author} {\bibfnamefont {J.~J.}\ \bibnamefont {Katz}},\
  }\href {\doibase 10.1016/0006-3002(63)90617-6} {\bibfield  {journal}
  {\bibinfo  {journal} {Biochimica et Biophysica Acta}\ }\textbf {\bibinfo
  {volume} {75}},\ \bibinfo {pages} {306} (\bibinfo {year} {1963})}\BibitemShut
  {NoStop}%
\bibitem [{\citenamefont {Du}\ \emph {et~al.}(1998)\citenamefont {Du},
  \citenamefont {Fuh}, \citenamefont {Li}, \citenamefont {Corkan},\ and\
  \citenamefont {Lindsey}}]{DuPhotochemCAD1998}%
  \BibitemOpen
  \bibfield  {author} {\bibinfo {author} {\bibfnamefont {H.}~\bibnamefont
  {Du}}, \bibinfo {author} {\bibfnamefont {R.-C.~A.}\ \bibnamefont {Fuh}},
  \bibinfo {author} {\bibfnamefont {J.}~\bibnamefont {Li}}, \bibinfo {author}
  {\bibfnamefont {L.~A.}\ \bibnamefont {Corkan}}, \ and\ \bibinfo {author}
  {\bibfnamefont {J.~S.}\ \bibnamefont {Lindsey}},\ }\href {\doibase
  10.1111/j.1751-1097.1998.tb02480.x} {\bibfield  {journal} {\bibinfo
  {journal} {Photochemistry and Photobiology}\ }\textbf {\bibinfo {volume}
  {68}},\ \bibinfo {pages} {141} (\bibinfo {year} {1998})}\BibitemShut
  {NoStop}%
\bibitem [{\citenamefont {Baldo}\ \emph {et~al.}(1999)\citenamefont {Baldo},
  \citenamefont {Lamansky}, \citenamefont {Burrows}, \citenamefont {Thompson},\
  and\ \citenamefont {Forrest}}]{irppy3_0}%
  \BibitemOpen
  \bibfield  {author} {\bibinfo {author} {\bibfnamefont {M.~A.}\ \bibnamefont
  {Baldo}}, \bibinfo {author} {\bibfnamefont {S.}~\bibnamefont {Lamansky}},
  \bibinfo {author} {\bibfnamefont {P.~E.}\ \bibnamefont {Burrows}}, \bibinfo
  {author} {\bibfnamefont {M.~E.}\ \bibnamefont {Thompson}}, \ and\ \bibinfo
  {author} {\bibfnamefont {S.~R.}\ \bibnamefont {Forrest}},\ }\href {\doibase
  http://dx.doi.org/10.1063/1.124258} {\bibfield  {journal} {\bibinfo
  {journal} {Appl. Phys. Lett.}\ }\textbf {\bibinfo {volume} {75}},\ \bibinfo
  {pages} {4} (\bibinfo {year} {1999})}\BibitemShut {NoStop}%
\bibitem [{\citenamefont {Fine}\ \emph {et~al.}(2012)\citenamefont {Fine},
  \citenamefont {Diri}, \citenamefont {Krylov}, \citenamefont {Nemirow},
  \citenamefont {Lu},\ and\ \citenamefont {Wittig}}]{FineIrppy3Spectrum2012}%
  \BibitemOpen
  \bibfield  {author} {\bibinfo {author} {\bibfnamefont {J.}~\bibnamefont
  {Fine}}, \bibinfo {author} {\bibfnamefont {K.}~\bibnamefont {Diri}}, \bibinfo
  {author} {\bibfnamefont {A.}~\bibnamefont {Krylov}}, \bibinfo {author}
  {\bibfnamefont {C.}~\bibnamefont {Nemirow}}, \bibinfo {author} {\bibfnamefont
  {Z.}~\bibnamefont {Lu}}, \ and\ \bibinfo {author} {\bibfnamefont
  {C.}~\bibnamefont {Wittig}},\ }\href {\doibase 10.1080/00268976.2012.685899}
  {\bibfield  {journal} {\bibinfo  {journal} {Molecular Physics}\ }\textbf
  {\bibinfo {volume} {110}},\ \bibinfo {pages} {1849} (\bibinfo {year}
  {2012})}\BibitemShut {NoStop}%
\bibitem [{\citenamefont {te~Velde}\ \emph {et~al.}(2001)\citenamefont
  {te~Velde}, \citenamefont {Bickelhaupt}, \citenamefont {Baerends},
  \citenamefont {Fonseca~Guerra}, \citenamefont {van Gisbergen}, \citenamefont
  {Snijders},\ and\ \citenamefont {Ziegler}}]{ADF2001}%
  \BibitemOpen
  \bibfield  {author} {\bibinfo {author} {\bibfnamefont {G.}~\bibnamefont
  {te~Velde}}, \bibinfo {author} {\bibfnamefont {F.~M.}\ \bibnamefont
  {Bickelhaupt}}, \bibinfo {author} {\bibfnamefont {E.~J.}\ \bibnamefont
  {Baerends}}, \bibinfo {author} {\bibfnamefont {C.}~\bibnamefont
  {Fonseca~Guerra}}, \bibinfo {author} {\bibfnamefont {S.~J.~A.}\ \bibnamefont
  {van Gisbergen}}, \bibinfo {author} {\bibfnamefont {J.~G.}\ \bibnamefont
  {Snijders}}, \ and\ \bibinfo {author} {\bibfnamefont {T.}~\bibnamefont
  {Ziegler}},\ }\href {\doibase 10.1002/jcc.1056} {\bibfield  {journal}
  {\bibinfo  {journal} {J. Comput. Chem.}\ }\textbf {\bibinfo {volume} {22}},\
  \bibinfo {pages} {931} (\bibinfo {year} {2001})}\BibitemShut {NoStop}%
\end{thebibliography}%

\begin{table*}[p]
\caption{\label{tab:parameters}Values for the magnetic Hubbard parameters~$W_\mathcal A$.}\vspace{10pt}
\begin{tabular}{c|c|c}
Element & $Z$ & $W_\mathcal{A}$ [Ha] \\\hline\hline
H  &  1 & -0.0717 \\\hline
He &  2 & -0.0865 \\\hline
Li &  3 & -0.0198 \\\hline
Be &  4 & -0.0230 \\\hline
B  &  5 & -0.0196 \\\hline
C  &  6 & -0.0226 \\\hline
N  &  7 & -0.0254 \\\hline
O  &  8 & -0.0278 \\\hline
F  &  9 & -0.0298 \\\hline
Ne & 10 & -0.0317 \\\hline
Na & 11 & -0.0152 \\\hline
Mg & 12 & -0.0166 \\\hline
Al & 13 & -0.0140 \\\hline
Si & 14 & -0.0144 \\\hline
P  & 15 & -0.0149 \\\hline
S  & 16 & -0.0155 \\\hline
Cl & 17 & -0.0161 \\\hline
Ar & 18 & -0.0166 \\\hline
K  & 19 & -0.0107 \\\hline
Ca & 20 & -0.0120 \\\hline
Sc & 21 & -0.0124 \\\hline
Ti & 22 & -0.0138 \\\hline
V  & 23 & -0.0141 \\\hline
Cr & 24 & -0.0138 \\\hline
Mn & 25 & -0.0150 \\\hline
Fe & 26 & -0.0154 \\\hline
Co & 27 & -0.0158 \\\hline
Ni & 28 & -0.0168 \\\hline
Cu & 29 & -0.0171 \\\hline
Zn & 30 & -0.0169 \\\hline
Ga & 31 & -0.0134 \\\hline
Ge & 32 & -0.0136 \\\hline
As & 33 & -0.0136 \\\hline
Se & 34 & -0.0137 \\\hline
Br & 35 & -0.0138 \\\hline
Kr & 36 & -0.0138 \\\hline
Rb & 37 & -0.0096 \\\hline
Sr & 38 & -0.0107 \\\hline
Y  & 39 & -0.0097 \\\hline
Zr & 40 & -0.0107 \\\hline
Nb & 41 & -0.0113 \\\hline
Mo & 42 & -0.0125 \\\hline
Tc & 43 & -0.0127 \\\hline
Ru & 44 & -0.0132 \\\hline
Rh & 45 & -0.0134
\end{tabular}
\hspace{1.2cm}
\begin{tabular}{c|c|c}
Element & $Z$ & $W_\mathcal{A}$ [Ha] \\\hline\hline
Pd & 46 & -0.0136 \\\hline
Ag & 47 & -0.0137 \\\hline
Cd & 48 & -0.0138 \\\hline
In & 49 & -0.0115 \\\hline
Sn & 50 & -0.0117 \\\hline
Sb & 51 & -0.0116 \\\hline
Te & 52 & -0.0115 \\\hline
I  & 53 & -0.0114 \\\hline
Xe & 54 & -0.0114 \\\hline
Cs & 55 & -0.0083 \\\hline
Ba & 56 & -0.0094 \\\hline
La & 57 & -0.0089 \\\hline
Ce & 58 & -0.0090 \\\hline
Pr & 59 & -0.0111 \\\hline
Nd & 60 & -0.0116 \\\hline
Pm & 61 & -0.0120 \\\hline
Sm & 62 & -0.0124 \\\hline
Eu & 63 & -0.0127 \\\hline
Gd & 64 & -0.0091 \\\hline
Tb & 65 & -0.0132 \\\hline
Dy & 66 & -0.0134 \\\hline
Ho & 67 & -0.0137 \\\hline
Er & 68 & -0.0139 \\\hline
Tm & 69 & -0.0141 \\\hline
Yb & 70 & -0.0142 \\\hline
Lu & 71 & -0.0090 \\\hline
Hf & 72 & -0.0098 \\\hline
Ta & 73 & -0.0104 \\\hline
W  & 74 & -0.0107 \\\hline
Re & 75 & -0.0109 \\\hline
Os & 76 & -0.0111 \\\hline
Ir & 77 & -0.0112 \\\hline
Pt & 78 & -0.0113 \\\hline
Au & 79 & -0.0108 \\\hline
Hg & 80 & -0.0114 \\\hline
Tl & 81 & -0.0107 \\\hline
Pb & 82 & -0.0110 \\\hline
Bi & 83 & -0.0109 \\\hline
Po & 84 & -0.0108 \\\hline
At & 85 & -0.0107 \\\hline
Rn & 86 & -0.0106 \\\hline
Fr & 87 & -0.0082 \\\hline
Ra & 88 & -0.0092 \\\hline
Ac & 89 & -0.0080 \\\hline
Th & 90 & -0.0084
\end{tabular}
\end{table*}

\end{document}